\documentclass[useAMS]{mn2e}

\usepackage{amssymb,amsmath,psfig,times, float}
\def\gsim{ \lower .75ex \hbox{$\sim$} \llap{\raise .27ex \hbox{$>$}} }
\def\lsim{ \lower .75ex\hbox{$\sim$} \llap{\raise .27ex \hbox{$<$}} }
\def\propsim{ \lower .75ex\hbox{$\sim$} \llap{\raise .27ex \hbox{$\propto$}} }

\def\ergsHz{{\rm\thinspace erg \thinspace s^{-1} \thinspace Hz^{-1}}}
\def\kms{{\rm\thinspace km \thinspace s^{-1}}}

\def\sc{Schwarzschild}
\def\beq{\begin{equation}}
\def\eeq{\end{equation}}

\def\sc{Schwarzschild}

\title[Jet--disc connection in AGN]
{The jet--disc connection in AGN}
\author[T. Sbarrato et al.]
{T. Sbarrato$^{1,2,3}$\thanks{Email: tullia.sbarrato@brera.inaf.it}, 
P. Padovani$^{1,4}$, G. Ghisellini$^2$
\\
$^1$ESO -- European Southern Observatory, Karl--Schwarzschild--Strasse 2, 
	D--85748 Garching bei M\"unchen, Germany\\
$^2$INAF -- Osservatorio Astronomico di Brera, Via Bianchi 46, I--23807 Merate, Italy\\
$^3$Universit\`a dell'Insubria, Dipartimento di Scienza e Alta Tecnologia, Via Valleggio 11, I--22100 Como, Italy\\
$^4$Associated to INAF -- Osservatorio Astronomico di Roma, via Frascati 33, I--00040 Monteporzio Catone, Italy\\
}

\begin{document}  

\maketitle

\begin{abstract}
We present our latest results on the connection between 
accretion rate and relativistic jet power in AGN,
by using a large sample which includes mostly blazars, but contains 
also some radio--galaxies. 
The jet power can be traced by $\gamma$--ray luminosity in the case of blazars, 
and radio luminosity for both classes. 
The accretion disc luminosity is instead traced by the broad emission lines. 
Among blazars, we find a correlation between broad line emission 
and the $\gamma$--ray or radio luminosities, suggesting a direct tight 
connection between jet power and accretion rate. 
We confirm that the observational differences between blazar subclasses 
reflect differences in the accretion regime, but with blazars only we cannot 
properly access the low--accretion regime. 
By introducing radio--galaxies, we 
succeed in observing the fingerprint of 
the transition between radiatively efficient and inefficient accretion discs 
in the jetted AGN family. 
The transition occurs at the standard critical value 
$L_{\rm d}/L_{\rm Edd}\sim10^{-2}$ and it appears smooth.  
Below this value, the ionizing luminosity emitted by the accretion structure drops significantly. 
\end{abstract}

\section{Introduction}

Blazars are Active Galactic Nuclei (AGN) with a relativistic jet directed towards our line of sight. 
Because of relativistic beaming, the emission from the jet is highly boosted, and dominates 
the AGN emission at all wavelengths, from the radio to the $\gamma$--ray band. 
They are very useful to study the relation between accretion 
structure and relativistic jet in AGN. 

Blazars are commonly divided in Flat Spectrum Radio Quasars (FSRQs) and BL Lacertae 
objects (BL Lacs), depending on the rest frame equivalent width (EW) of their broad emission lines. 
Specifically, if the EW is smaller than 5\AA, the object is classified as a BL Lac, otherwise 
as a FSRQ (Urry \& Padovani 1995). 
An analogous classification criterion was introduced by Landt et al.\ (2004), that found  
possible to discriminate between objects with intrinsically weak or strong narrow emission lines 
by studying the [OII] and [OIII] EW plane. 
The EW--based classification, besides being simple to apply, is based on the fact that  
the EW is a good measure of the line emission dominance over the non--thermal continuum emitted 
from the jet. 
In this view, the EW could tell if an object had intrinsically strong (FSRQ) or weak (BL Lac) emission lines. 
However, jet emission is extremely variable, definitely more than the thermal continuum 
and the emission lines. 
Hence the line EW can dramatically vary from one state to another for the same source. 
A blazar with intrinsically very luminous emission lines can temporarily appear as a BL Lac, with very small 
EW, if its jet flux happens to be more luminous than usual. 
On the contrary, during a particularly low state, a BL Lac can show emission lines with EW larger 
than the 5\AA\ limit (as it happened to BL Lac itself; Vermeulen et al.\ 1995; Capetti, Raiteri \& Buttiglione 2010). 
Ghisellini et al.\ 2011 (hereafter G11) already 
proposed that this classification is not reliable, and moreover does not reflect any intrinsic 
property or difference within the blazar class, being dependent on the strong continuum variability. 
These authors introduced a more physical classification, based on the different accretion rates of the 
two subclasses of blazars, later confirmed in our earlier work (Sbarrato et al.\ 2012; hereafter TS12): 
FSRQs have a disc luminosity $\gsim5\times10^{-3}$ of the Eddington one. 
In those papers,  a strong correlation between accretion and jet emissions was found.  
The unreliability of the EW classification was also deeply investigated by Giommi et al.\ (2012; 2013), 
who suggested that blazars should be divided in high-- and low--ionization 
sources, instead of focusing on observed features that are not physically relevant. 
In other words, G11, TS12 and Giommi et al.\ (2012; 2013) based their reclassification on the 
ionizing luminosity, emitted from the accretion disc. 

Radio--galaxies are thought to be the parent population of blazars: 
while blazars are aligned to our line--of--sight, radio--galaxies have their 
jets oriented at larger viewing angles. 
This greatly affects the overall emission and the typical Spectral Energy Distribution (SED), 
that is not dominated by the non--thermal radiation in the case of radio--galaxies, 
contrary to blazars (Urry \& Padovani 1995).
Radio--galaxies are historically divided into two subclasses according to their radio morphology 
(Fanaroff \& Riley 1974): FRI show bright jets close to the nucleus, while FRII show 
prominent hot spots far from it. 
This classification also reflects in a separation in their radio power 
(below and above $L_{\rm 178 MHz}=2.5\times10^{33}\ergsHz$, respectively). 
As in the case of blazars, the classification is not sharp, but radio--galaxies 
seem to be distributed continuously between the two classes. 
Similarly to the blazar case, radio--galaxies were also classified according to their 
ionization efficiency. 
In the nineties many spectroscopic--based classifications were proposed, also 
connecting the emission line luminosities to the radio power 
(e.g.\ Baum \& Heckman 1989a,b; Rawlings et al.\ 1989; Rawlings \& Saunders 1991; 
Morganti et al.\ 1997; Labiano 2008; Willott et al.\ 1999). 
Ghisellini \& Celotti (2001) showed that the division between FRI and FRII actually 
reflected a systematic difference in accretion rate: FRI were shown to have generally 
an ionizing luminosity $\sim10^{-2}-10^{-3}$ of the Eddington one, while for FRII this was
typically larger. 
Laing et al.\ (1994) introduced a sub--classification of FRII sources into high--excitation 
(HEG) and low--excitation galaxies (LEG), but this was found to be applicable also to 
some FRI. 
Therefore, Buttiglione et al.\ (2009; 2010) decided to perform an extended investigation 
of the spectroscopic properties of radio--galaxies, using a homogeneous sample. 
They found that all HEG are FRII, while LEG can be both FRI and FRII. 
While a ionization--based classification seems to be more physically relevant also in the 
case of radio--galaxies, an univocal, physically based classification method is not 
yet commonly accepted, as in the case of blazars. 

In this work, we try to investigate the relation between accretion and jet emissions 
in jetted AGN, to understand if a change in the accretion mode happens inside 
the blazar family, and the whole jetted AGN class. 
We enlarge the blazar sample with respect to the one in TS12, and we extend our study 
to another tracer of the jet power: radio luminosity. 
In this way, we can also include radio--galaxies to study the jet--accretion relation 
in the whole family of jetted AGN. 
Section \ref{sample} presents the samples, Section \ref{blr} describes how we trace the accretion 
luminosity through the broad line region emission. 
In Section \ref{gamma} we investigate the relationship between broad line region and 
$\gamma$--ray power, while the one between broad line region and radio power
is dealt with in Section \ref{radio}. 
The discussion is presented in Section \ref{discussion}, and our results are summarized 
in Section \ref{conclusion}.

\section{The sample}
\label{sample}

We are interested in studying the relationship between jet and accretion in AGN, 
through $\gamma$--ray or radio luminosity and broad line region (BLR) luminosity, respectively. 
We first tried to study it among the blazar subclass, by collecting a complete sample of 
$\gamma$--ray detected blazars, with measurements of the luminosity of broad lines  
obtained through optical spectroscopy. 
We then extended our sample to radio--galaxies intrinsically without broad emission lines (but with known redshift), 
to investigate the jet--accretion relationship in the inefficient accretion regime. 
Even if not complete, the radio--galaxy sample we describe in \S\ref{sample_rg} is the 
most useful for our purposes, as the ionization status of every member is throughly studied. 

\subsection{The blazar samples}

The Large Area Telescope (LAT) onboard the {\it Fermi Gamma-Ray Space Telescope} ({\it Fermi}) 
has detected a large amount of blazars in the $\gamma$--rays.  
In two different papers (Shaw et al.\ 2012 and Shaw et al.\ 2013; respectively S12 and S13 hereafter), 
Shaw and collaborators obtained optical spectroscopic data for all the FSRQs (S12) and most of the 
BL Lacs (S13) included in the Second Catalog of AGN Detected by the Fermi LAT (2LAC; Ackermann et al.\ 2011). 
We collected from these two papers all the blazars that show broad emission lines in their optical spectra.

In S12, the authors analyzed the optical spectra of 229 FSRQs: they obtained new spectra for 
165 FSRQs included in the First Catalog of AGN detected by {\it Fermi} (1LAC; Abdo et al.\ 2010), 
and re--analyzed Sloan Digital Sky Survey (SDSS) spectra of  64 other FSRQs (not all with 
$\gamma$--ray data). 
Along with spectroscopic data, the authors derived virial black hole masses for all their objects. 
In order to have a complete description of the sources, we selected only the FSRQs with enough data to 
fit the entire SED.
We are left with 191 objects. 
We cross--correlated this sample with the 2LAC. 
When the objects were not included in the 2LAC, we collected data from the 1LAC. 
The sample was completed including radio data from the Combined Radio All-Sky Targeted 
Eight GHz Survey (CRATES; Healey et al.\ 2007). 
When sources were not included in CRATES (only 5 objects), we obtained radio data 
at frequencies close to 8~GHz from the ASI Science Data Center (ASDC). 
We are then left with a sample of 180 FSRQs with optical spectroscopy, along with $\gamma$--ray and 
radio data.

In S13, instead, the BL Lac objects are directly selected from the 2LAC sample. 
2LAC includes 475 BL Lacs, among which only 209 have redshift information. 
We consider here only those objects with broad emission lines, in order to be able to derive 
an estimate of $L_{\rm BLR}$, avoiding upper limits. 
We are left with 26 objects. 
All of them have 2LAC $\gamma$--ray and CRATES radio data. 
S13 did not derive the black hole masses for these objects. 
We can anyway assign an average $M_{\rm BH}$ value to these BL Lacs, 
from an adapted version of the $M_{\rm BH}-M_R$ relation (from Bettoni et al.\ 2003), 
that assumes a mass ratio between bulge and central black hole of $10^3$. 
In fact, BL Lacs are typically hosted in massive and luminous elliptical galaxies, 
with a small dispersion in absolute magnitude 
($\langle M_R\rangle=-22.8\pm0.5$; Sbarufatti, Treves \& Falomo 2005). 
Therefore, we can derive an average value of their black hole mass: 
\begin{equation}
\log\left(\frac{M_{\rm BH}}{M_\odot}\right) \simeq -0.5\times M_R -3 = 8.4.
\end{equation}
Since we need the central black hole masses for our studies, we will apply this 
value to every object without a mass estimate. 
These BL Lacs and their relevant data are listed in Table \ref{s13_bllac}. 

We add to these new samples the objects that show broad emission lines studied in TS12. 
Specifically, we add the 45 FSRQs and 1 BL Lac (following our reclassification) that 
were included in Shen et al.\ (2011), along with the 15 BL Lacs with broad emission lines 
from Ghisellini et al.\ (2011).
In this work, we only consider the objects from TS12 that have broad emission lines detected, 
excluding then all the BL Lacs with only an upper limit on the BLR luminosity. 
In TS12, the upper limits were introduced to increase the number of BL Lacs, 
populating the low--luminosity branch of our sample. 
They followed the $L_{\rm BLR}-L_\gamma$ correlation already found only with the 
detections, therefore they were not very constraining. 
In our new work, we increased the number of BL Lacs with broad emission lines thanks to S13, 
and therefore we do not need the lineless BL Lacs. 

In total, we have 225 FSRQs and 42 BL Lacs with detected broad emission lines, 
both $\gamma$--ray and radio counterparts, and a reliable estimate of the central black hole 
masses ($M_{\rm BH}$). 

\begin{table*} 
\centering
\begin{tabular}{lccccccccccc|}
\hline
\hline
Name & Fermi Name & RA & DEC & $z$ & Line & $\log L_{\rm BLR}$ & $\log L_\gamma$ & $\log L_{\rm radio}$ \\ 
~[1] &[2] &[3] &[4] &[5] &[6] &[7] &[8] &[9] \\
\hline
{\bf BL Lac} &	&	&	&	&	&	&	&	\\
 GB6 J0013+1910 & 2FGLJ0013.8+1907 & 00 13 56.3 & +19 10 41.5 & 0.477 & MgII & 42.691 & 45.441 & 43.043\\
 PKS 0829+046 & 2FGLJ0831.9+0429 & 08 31 48.7 & +04 29 38.2 & 0.174 & Ha & 42.614 & 45.520 & 42.923\\
 RBS 0958 & 2FGLJ1117.2+2013 & 11 17 06.1 & +20 14 07.6 & 0.138 & Ha & 41.722 & 44.723 & 41.622\\
 PMN J1125-3556 & 2FGLJ1125.6-3559 & 11 25 31.3 & -35 57 05.0 & 0.284 & Ha & 43.338 & 45.157 & 42.627\\
 SBS 1200+608 & 2FGLJ1203.2+6030 & 12 03 03.4 & +60 31 19.1 & 0.065 & Ha & 42.005 & 43.769 & 41.190\\
 W Comae & 2FGLJ1221.4+2814 & 12 21 31.6 & +28 13 58.1 & 0.103 & Ha & 42.137 & 45.055 & 42.199\\
 PG 1218+304 & 2FGLJ1221.3+3010 & 12 21 21.9 & +30 10 36.2 & 0.184 & Ha & 42.063 & 45.174 & 41.734\\
 OQ 530 & 2FGLJ1420.2+5422 & 14 19 46.5 & +54 23 15.0 & 0.153 & Ha & 42.201 & 44.766 & 42.603\\
 RGB J1534+372 & 2FGLJ1535.4+3720 & 15 34 47.2 & +37 15 53.8 & 0.144 & Ha & 41.722 & 44.349 & 40.991\\
\hline
{\bf BL/FS} &	&	&	&	&	&	&	&	\\
 PKS 0332-403 & 2FGLJ0334.2-4008 & 03 34 13.4 & -40 08 26.9 & 1.357 & MgII & 45.046 & 47.711 & 44.967\\
 TXS 0431-203 & 2FGLJ0434.1-2014 & 04 34 07.9 & -20 15 17.2 & 0.928 & MgII & 43.153 & 46.511 & 43.758\\
 PKS 0437-454 & 2FGLJ0438.8-4521 & 04 39 00.7 & -45 22 23.9 & 2.017 & CIV & 45.201 & 47.661 & 45.363\\
 PKS 0627-199 & 2FGLJ0629.3-2001 & 06 29 23.7 & -19 59 19.7 & 1.724 & CIV & 44.060 & 47.740 & 45.036\\
 4C +14.60 & 2FGLJ1540.4+1438 & 15 40 49.5 & +14 47 46.5 & 0.606 & MgII & 43.575 & 45.966 & 44.229\\
 PMN J1754-6423 & 2FGLJ1755.5-6423 & 17 54 41.8 & -64 23 44.7 & 1.255 & MgII & 44.163 & 46.912 & 44.113\\
 4C +56.27 & 2FGLJ1824.0+5650 & 18 24 07.0 & +56 51 01.1 & 0.664 & MgII & 43.912 & 46.891 & 44.333\\
 S3 2150+17 & 2FGLJ2152.4+1735 & 21 52 24.7 & +17 34 37.9 & 0.874 & MgII & 44.168 & 46.333 & 44.310\\
 PMN J2206-0031 & 2FGLJ2206.6-0029 & 22 06 43.2 & -00 31 02.3 & 1.053 & MgII & 43.798 & 46.557 & 43.858\\
 B2 2234+28A & 2FGLJ2236.4+2828 & 22 36 22.3 & +28 28 58.1 & 0.79 & MgII & 44.645 & 47.079 & 44.412\\
 PKS 2244-002 & 2FGLJ2247.2-0002 & 22 47 30.1 & +00 00 07.0 & 0.949 & MgII & 44.106 & 46.543 & 43.950\\
 PKS 2312-505 & 2FGLJ2315.7-5014 & 23 15 44.2 & -50 18 39.7 & 0.811 & MgII & 43.628 & 46.357 & 43.764\\
 PKS 2351-309 & 2FGLJ2353.5-3034 & 23 53 47.3 & -30 37 48.3 & 0.737 & MgII & 43.628 & 46.066 & 43.893\\
\hline
{\bf FS} &	&	&	&	&	&	&	&	\\
 NVSS J020344+304238 & 2FGLJ0204.0+3045 & 02 03 44.1 & +30 42 38.1 & 0.761 & MgII & 44.757 & 46.449 & 43.572\\
 PKS 0516-621 & 2FGLJ0516.8-6207 & 05 16 44.5 & -62 07 04.8 & 1.3 & MgII & 44.444 & 47.488 & 44.523\\
 MG2 J201534+3710 & 2FGLJ2015.6+3709 & 20 15 28.6 & +37 10 59.8 & 0.859 & Hb & 44.342 & 47.971 & 44.794\\
 TXS 2206+650 & 2FGLJ2206.6+6500 & 22 08 03.3 & +65 19 38.7 & 1.121 & MgII & 44.336 & 47.324 & 44.360\\
\hline
\hline
\end{tabular}
\vskip 0.4 true cm
\caption{Sources from the S13 BL Lac sample, divided according to their reclassification (see \S\ref{gamma}).
Col. [1]: name;
Col. [2]: {\it Fermi}/LAT counterpart;
Col. [3]: right ascension;
Col. [4]: declination;
Col. [5]: redshift;
Col. [6]: line measured, from which $L_{\rm BLR}$ has been derived;
Col. [7]: logarithm of the broad line region luminosity (in erg s$^{-1}$); 
Col. [8]: logarithm of the $\gamma$--ray luminosity from {\it Fermi} data (in erg s$^{-1}$);
Col. [9]: logarithm of the radio luminosity, calculated at 8~GHz rest--frame (in erg s$^{-1}$).
}
\label{s13_bllac}
\end{table*}

\subsection{The radio--galaxy sample}
\label{sample_rg}

We collected a sample of radio--galaxies without broad emission lines from the work by Buttiglione et al.\ (2010). 
The authors studied the optical spectroscopical and radio features of all the $z<0.3$ 
radio--galaxies (Buttiglione et al.\ 2009) with $F_{\rm 178MHz}>9{\rm Jy}$, $\delta>-5^\circ$ 
and an optical counterpart from the Third Cambridge radio 
catalogue (3CR; Spinrad et al.\ 1985).
The authors classify the sources in high--excitation (HEG), low--excitation galaxies (LEG) and 
broad line objects (BLO) according to optical features. 
Specifically, BLOs clearly show broad emission lines, while 
HEGs and LEGs do not show any broad emission feature, but while the latter have 
an intrinsic lack of broad line emitting structures, the former show high--excitation 
fingerprints, suggesting obscuration of the broad line region more than a true absence. 
The introduction of radio--galaxies in our work aims at studying the true lineless jetted AGN, 
so we include in our sample the 37 LEGs studied by Buttiglione et al.\ (2010). 

Along with the optical spectral analysis, they performed a radio morphology study. 
The radio analysis provides information on the core power, which allows us to have a reliable tracer 
of the inner jet power, without contamination from the extended structures.
In other words, the core power traces the same emission that is traced 
by the radio luminosity in the case of blazars, apart from the 
different beaming factor (due to the different orientation angles of the jet, see \S\ref{radio}). 
Moreover, the authors provide a measure of the $H$ magnitude, and a calibration to 
obtain from that the central black hole mass of the objects (Marconi \& Hunt 2003): 
\begin{equation}
\log\left(\frac{M_{\rm BH}}{M_\odot}\right) = -2.77- 0.464\times M_H.
\end{equation} 
The authors find different radio morphologies among the 37 LEGs: 
12 FRI, 16 FRII and 9 unclassifiable sources. 
One object out of each radio group has no $M_H$ information, so we cannot derive a black 
hole mass estimate. 
We exclude from our sample those objects. 

We are then left with 11 LEGs FRI, 15 LEGs FRII and 8 LEGs without a radio classification (FR?), 
all with an estimate of the radio core power, with narrow emission line information that give upper 
limits on the broad emission lines (see \S\ref{blr}), and black hole mass estimates. 
We also added to the radio--galaxy sample M87, a ``classic" LEG FRI, taking the radio flux from NED, 
while the optical spectroscopic information is taken from Buttiglione et al.\ (2009). 
M87 does not show any broad emission line, so in the following we will treat it 
as the LEGs studied by Buttiglione et al.\ (2010).

\section{The BLR Luminosity}
\label{blr}

To calculate the BLR luminosity, we need an estimate of the broad emission lines of each object, 
specifically of Ly$\alpha$, H$\alpha$, H$\beta$, MgII or CIV. 
In the case of the S12 FSRQs, we take directly the emission lines fluxes listed in S12, as we did in TS12, 
while in the case of BL Lacs we directly fit the few broad emission lines present. 
G11 had already derived from literature the broad emission line values for the BL Lacs included in their sample. 
Note that the emission lines visible in Mkn421 and Mkn501 spectra have a FWHM ($\sim1000\kms$) 
that lies on the classification threshold between narrow lines and broad lines. 
Therefore, we rather put an upper limit on their broad line luminosities, to be more conservative. 
We nevertheless leave them in our sample, since they are usually listed as 
known BL Lacs with lines. 
The objects included in our radio--galaxy sample have the same issue, being explicitly selected 
as sources intrinsically lacking broad emission lines.  
They only show narrow emission lines, so we can obtain from them upper limits on  
broad emission line luminosities, taking as example quasars with both broad 
and narrow emission lines. 
Narrow line--dominated (but with broad emission lines) quasars are in fact included in S11. 
In their case, the ratio between narrow and broad H$\alpha$ luminosities is: 
\begin{equation}
\frac{L_{\rm nH\alpha}}{L_{\rm bH\alpha}}\lsim10.
\end{equation}
In other words, if present, the broad emission lines have a luminosity that is at least 10\% 
the corresponding narrow line luminosity, or more. 
To be conservative, we choose to fix as a robust upper limit on the broad line luminosity, 
the luminosity associated to the corresponding detected narrow emission line.

From the broad emission line luminosities or from the upper limits collected 
for our sample, we can then derive the overall luminosity emitted from the BLR. 
We follow Celotti et al. (1997) and set 
 the Ly$\alpha$ flux contribution to 100, and the relative weights
of the H$\alpha$, H$\beta$, MgII and CIV lines to 77, 22, 34 and 63, respectively
(see Francis et al. 1991). The total broad line flux is fixed at 555.76.
The $L_{\rm BLR}$ value of each source has been derived
using these proportions. When more than one line is
present, we calculate the logarithmic average of the $L_{\rm BLR}$ estimated from
each line.

\subsection{$L_{\rm BLR}$ as a tracer of the accretion}
\label{blr_accr}

The BLR luminosity is a reliable tracer of the emission from the accretion structure, 
since the plasma producing the broad emission lines is directly ionized by its radiation. 
The fraction of disc luminosity that ionizes the plasma in the BLR (photoionizing luminosity, $L_{\rm ion}$) 
depends on the geometry of the disc, and hence on its radiation efficiency. 
In the simplest hypothesis for a radiatively efficient accretion disc (e.g.\ Shakura \& Sunyaev 1973), 
the disc is geometrically thin and emits as a blackbody at all radii. 
For simplicity, we approximate the photoionizing luminosity with the entire disc luminosity 
($L_{\rm d}=\eta\dot Mc^2$). 
Therefore we can assume that $L_{\rm ion}$ follows the same dependence on the accretion rate 
as the disc luminosity: 
\begin{equation} 
L_{\rm ion}\sim L_{\rm d}\propto\dot M.
\end{equation}
In this case, then, assuming an average covering factor of $\sim10\%$  for the BLR, 
the disc luminosity is $L_{\rm d}\sim 10L_{\rm BLR}$
(e.g. Baldwin \& Netzer 1978; Smith et al. 1981). 
Following the Shakura \& Sunyaev hypothesis, 
such a radiatively efficient disc should occur for $\dot m=\dot M/\dot M_{\rm Edd}>\dot m_c$, 
where $\dot M_{\rm Edd}=L_{\rm Edd}/c^2$ and  $\dot m_c\sim0.1$  
(Narayan \& Yi 1995). 
Sharma et al.\ (2007) suggested instead that the disc remains radiatively efficient 
for $\dot m_c\gsim 10^{-4}$. 

For values $\dot m<\dot m_c$ the accretion structure should become 
radiatively inefficient,  
switching from a proper accretion disc to a hot accretion flow. 
In this case, it does not emit as a blackbody, and the overall 
luminosity decreases and changes its dependence on the accretion 
rate to $L_{\rm d}\propto\dot M^2$ (Narayan et al. 1997).
As a consequence of the change in the emission mode, the ionizing luminosity 
is no more a relevant fraction of the overall disc luminosity, but $L_{\rm ion}\ll L_{\rm d}$. 
The accretion flow cannot emit as a blackbody, because it is no longer optically thick, 
and the emission is dominated by synchrotron, bremsstrahlung and comptonization 
processes (see Yuan \& Narayan 2014 for a review). 
The SED of the accretion flow is therefore very different compared to a 
Shakura--Sunyaev--like accretion disc (see Fig.\ 1 in Mahadevan 1997). 
The photoionizing luminosity (mostly the luminosity emitted in the UV wavelength range) 
is no more a fixed fraction of the disc luminosity. 
Since this fraction decreases when the accretion rate decreases, 
the photoionizing luminosity scales with the accretion rate following a relation 
steeper than $\propto\dot M^2$.
Following the model in Mahadevan (1997), in TS12 we estimated the ionizing luminosity 
for the different accretion rates considered by the author. 
The UV wavelength range is the most affected from the change in accretion 
rate of the whole SED, and in fact we found a very steep relation: 
\begin{equation}
L_{\rm ion}\propto\dot M^{3.5}.
\end{equation}
This strong change would lead to an analogous change in the dependence of $L_{\rm BLR}$ 
on $\dot M$. 

\section{The $L_{\rm BLR}$--$L_\gamma$ relation}
\label{gamma}

\begin{figure*}
\centering
\vskip -0.6 cm 
\hskip -0.4 cm
\psfig{figure=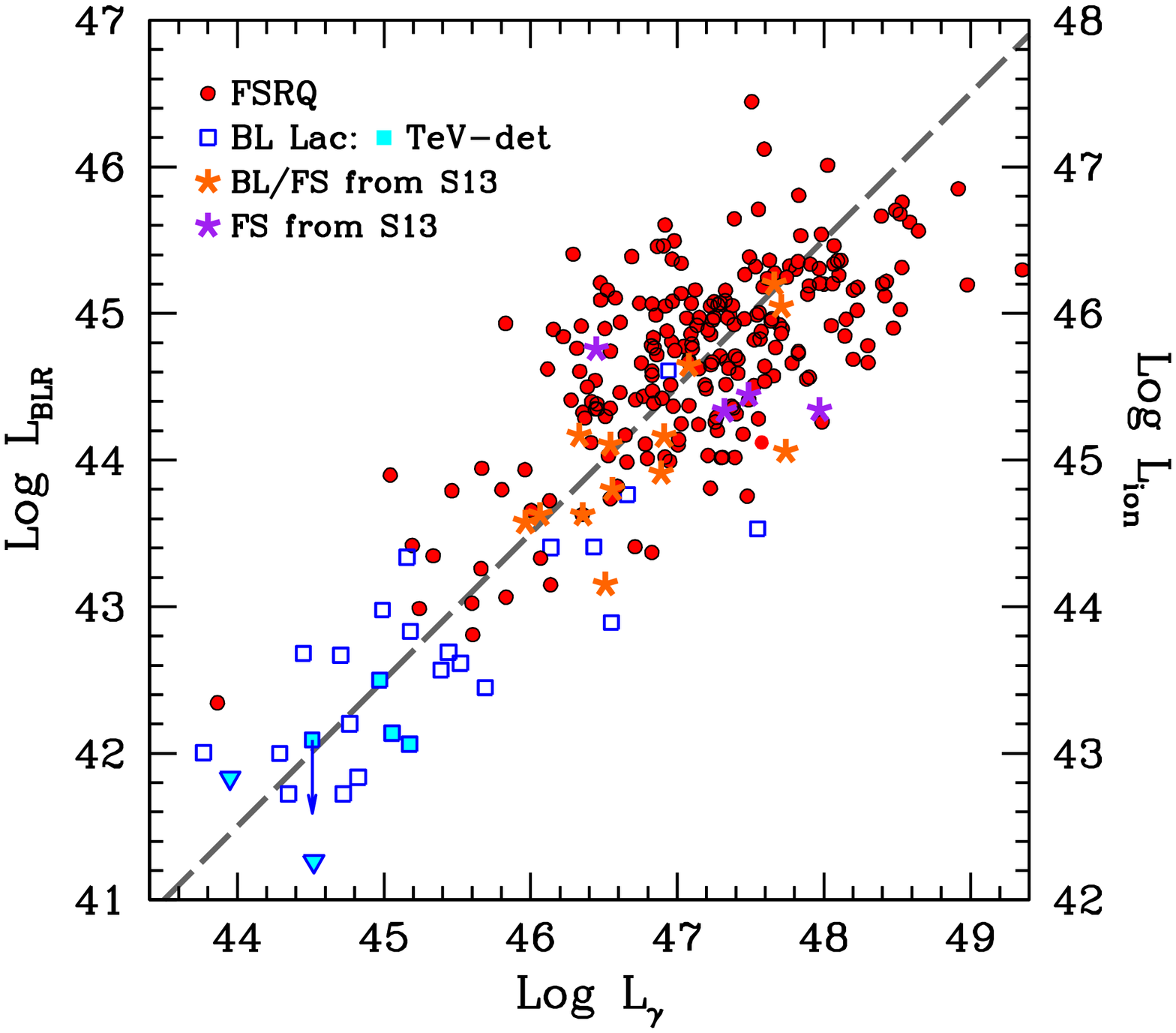,width=9cm,height=9cm}
\psfig{figure=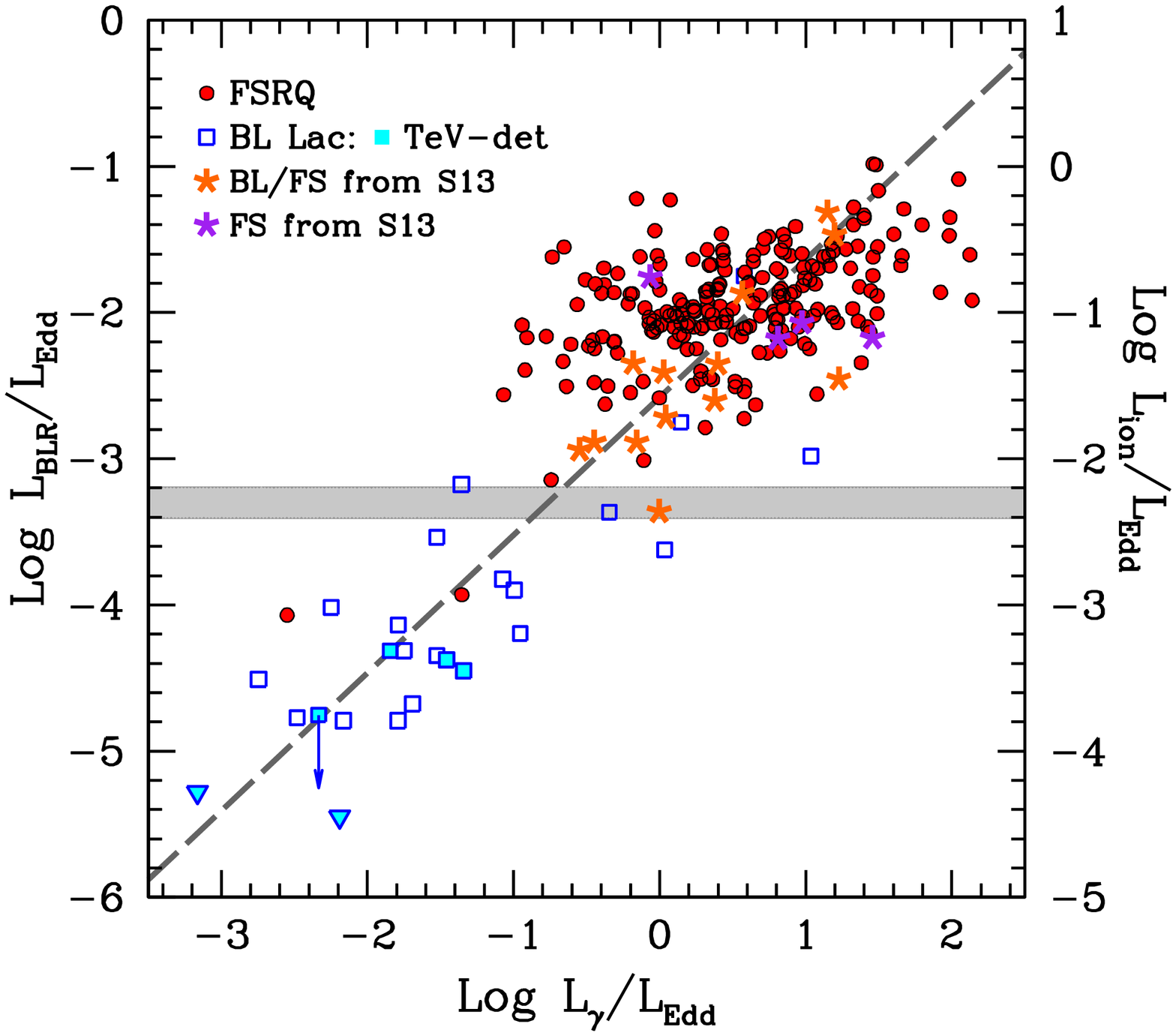,width=9cm,height=9cm}
\vskip -0.5 cm
\caption{
Broad line region luminosity as a function of $\gamma$--ray luminosity (left panel) 
and the same, normalized at the Eddington luminosity (right panel). 
Different symbols correspond to different classifications, as labelled. 
The dashed lines are the results of the least square fits described in Table \ref{correlation}. 
The grey stripe in the right panel indicates the luminosity divide between FSRQs and 
BL Lacs, located at $L_{\rm BLR}/L_{\rm Edd}\sim5\times10^{-4}$.
}
\label{blr_gamma}
\end{figure*}

\begin{table} 
\centering
\begin{tabular}{l l l l l l l }
\hline
\hline
    &$m$ &$q$ &$r$\\ 
\hline   
\multicolumn{3}{l}{$x = \log L_\gamma$; $y = \log  L_{\rm BLR}$} & \\	
$(x,y)$               &0.92$\pm$0.19 &1.2$\pm$14.8 &0.81\\ 
$(x,y)$, $z$       &0.92$\pm$0.19 &1.2$\pm$14.8 &0.59\\ 
\hline
\multicolumn{3}{l}{$x = \log (L_\gamma/L_{\rm Edd})$; $y = \log (L_{\rm BLR}/L_{\rm Edd})$} & \\ 
$(x,y)$       	    &0.84$\pm$0.20 &--2.46$\pm$3.1 &0.78\\ 
$(x,y)$, $z$          &0.84$\pm$0.20 &--2.46$\pm$3.1 &0.65\\ 
$(x,y)$, $z$, $M$ &0.84$\pm$0.20 &--2.46$\pm$3.1 &0.64\\ 
\hline
\hline 
\end{tabular}
\vskip 0.4 true cm
\caption{
Results of the partial correlation analysis of the $L_{\rm BLR}-L_{\gamma}$ and 
$L_{\rm BLR}/L_{\rm Edd}-L_\gamma/L_{\rm Edd}$ relations 
using a least square fit. The whole blazar sample is taken into account, i.e.\ 267 objects 
are considered in the analysis.
Correlations are of the form $y = mx +q$. 
The listed slopes $m$ refer to the bisector (of the two correlations $x$ vs $y$ and $y$ vs $x$).
$r$ is the correlation coefficient obtained from the analysis.
We list also the results when accounting for the common dependence on redshift and/or 
black hole mass.
In all the cases, the probability that the correlation is random is $P<4\times10^{-8}$, 
i.e.\ all the correlations are statistically relevant. 
}
\label{correlation}
\end{table}

Figure \ref{blr_gamma} represents the first result of our work. 
The left panel shows the 
BLR luminosity as a function of the $\gamma$--ray luminosity. 
The right panel shows the same quantities divided by the 
Eddington luminosity ($L_{\rm Edd}$). 
The objects included in TS12 are marked as FSRQs or BL Lacs, according 
to how we classified them in our previous work. 
The new FSRQs from S12 are included as FSRQs, while the objects classified 
as BL Lacs in S13 are marked as BL Lacs, BL/FS or FS in the plots. 
Note that the correlation we found in TS12 is confirmed by the new data, 
both when directly comparing the two luminosities and when 
normalizing them by the Eddington luminosity. 
This clearly strengthens the hypothesis of a tight relation between 
the accretion rate and the jet power in blazars. 
As explained in \S\ref{blr_accr}, the $L_{\rm BLR}$ is a very good tracer 
of the accretion rate, while the $\gamma$--ray luminosity traces well 
the jet power. 
We calculate the best fit of the relation between the two luminosities, 
both normalized by the Eddington luminosity and not. 
We find that both are consistent with the results found in TS12. 
We apply a partial correlation analysis, to take also into account 
the possible common dependence on $z$ and $M_{\rm BH}$ of the 
values. The $L_{\rm BLR}-L_{\gamma}$ and 
$L_{\rm BLR}/L_{\rm Edd}-L_\gamma/L_{\rm Edd}$ relations result 
linear and statistically relevant (see Table \ref{correlation}). 

Contrary to our previous work, instead, the apparent `divide' between 
FSRQs and BL Lacs (located at $L_{\rm BLR}/L_{\rm Edd}\sim5\times10^{-4}$) 
seems no longer valid, since some BL Lacs from S13 are located in the 
high--luminosity branch of the correlation, in the area typically occupied by FSRQs 
(S13 BL Lacs are marked as purple and orange asterisks, or included 
among the blue empty squares in all the Figures). 
To understand this discrepancy with our previous results, we first inspected visually 
the overall SEDs of the BL Lacs from S13. 
We notice that the sources show three different SED behavior 
(as shown by the individual SEDs in the Appendix):
\begin{itemize}

\item
nine have the synchrotron emission dominant or comparable to the high--energy 
component, and the thermal emission from the accretion structure completely 
swamped by the non--thermal jet emission. 
These features define a BL Lac, according to the 
classification scheme adopted in G11 and TS12, 
and first introduced by Padovani \& Giommi (1995). 

\item 
four of them show a clear Compton dominance, and the emission from the 
accretion disc is clearly visible. 
We then classify them as FSRQs, and claim for a misclassification in S13. 
However, we highlight them differently and label them as `FS', to keep track of them in the plots. 

\item
thirteen objects have the high--energy component that slightly dominates the synchrotron emission, 
as in the case of non extreme FSRQs. On the other hand, 
the synchrotron component completely swamps the accretion emission, leading 
to a BL Lac--like optical appearance.  
We classify them as `BL/FS', since they show both a FSRQ and a BL Lac fingerprint. 
\end{itemize}
The objects classified as FS and FS/BL are labelled accordingly in all our Figures. 
From Figure \ref{blr_gamma}, we immediately notice that all these ``reclassified" BL Lacs 
are the S13 BL Lacs that occupy the high--luminosity branch of our correlations. 
The FS have all the typical FSRQ features, so we expect to find them in the high--luminosity 
branch of the $L_{\rm BLR}-L_\gamma$ correlations (see Fig. \ref{blr_gamma}). 
Interestingly, all the others objects from S13 that were located in the FSRQ branch 
are the 13 that we classified as BL/FS. 
Their location allows us to better understand their peculiar SED features. 
We can in fact infer that they have an intrinsic powerful jet and a highly luminous 
accretion disc (i.e.\ high accretion rate), 
as common FSRQs, even if their optical spectroscopical features are BL Lac--like. 
In other words, the BL/FS were classified as BL Lacs because of an 
unusually powerful synchrotron emission that reduced the EW of their 
broad emission lines, but are instead FSRQs. 
Even a very luminous thermal continuum, with the related emission features, 
can in fact be overcome by a very luminous non--thermal continuum 
(Giommi et al.\ 2012; 2013). 
Since the synchrotron emission is mainly driven by the energy density of the magnetic field, 
we can expect that these objects have it unusually high. 
This is confirmed by the SED modeling. 
We fitted the overall SEDs with a one--zone leptonic model (Ghisellini \& Tavecchio 2009), 
and the results show that the energy density of the magnetic field of all 
the BL/FS is unusually high (detailed results will be shown in Ghisellini et al., in preparation). 

Considering the reclassification, the division between FSRQs and BL Lacs at 
$L_{\rm BLR}/L_{\rm Edd}\sim5\times10^{-4}$ becomes even more relevant. 
The BL/FS are actually FSRQs ``disguised" as BL Lacs and the canonical classification 
based on the equivalent width of their broad emission lines fails in classifying them.  
The division based on $L_{\rm BLR}/L_{\rm Edd}$ represents a more physical 
classifying system, since it discriminates the objects in terms of their accretion rate. 
However, the divide is not sharp, and again our blazar sample seems to be distributed 
continuously in both the  $L_{\rm BLR}-L_\gamma$ and the 
 $L_{\rm BLR}/L_{\rm Edd}-L_\gamma/L_{\rm Edd}$ planes. 
 
Along with the divide, we are interested in studying at what accretion rate (and if) 
a change in the accretion structure happens. 
As detailed in \S\ref{blr_accr}, a standard Shakura--Sunyaev disc should occur  
for accretion rates larger than a critical value $\dot m_c$. 
Below that value, the accretion structure is no longer radiatively efficient, also 
ionizing less efficiently the plasma in the broad line region. 
This change in accretion should hence be reflected in a change of slope 
in the $L_{\rm BLR}/L_{\rm Edd}-L_\gamma/L_{\rm Edd}$ plot. 
The jet power is in fact directly correlated to the accretion rate at all 
values of the accretion rate itself
(Celotti \& Ghisellini 2008; Ghisellini et al.\ 2010).
In TS12, the low--luminosity branch of the  
$L_{\rm BLR}/L_{\rm Edd}-L_\gamma/L_{\rm Edd}$ plot was not populated 
enough to draw a firm conclusion. 
The new BL Lacs have increased the number of objects that could help in 
understanding the possible existence of a break in the relation, but the data 
are still too sparse to draw a firm conclusion. 
Hence, we try to have a new perspective on the problem, by introducing another 
tracer for the jet power, that allows us to reach smaller accretion rates 
and observe directly the behavior of the jet--disc system in the case of 
truly inefficient accretion structures, i.e.\ objects intrinsically without broad emission lines.

\section{The $L_{\rm BLR}$--$L_{\rm radio}$ relation}
\label{radio}

\begin{figure}
\centering
\vskip -0.6 cm 
\psfig{figure=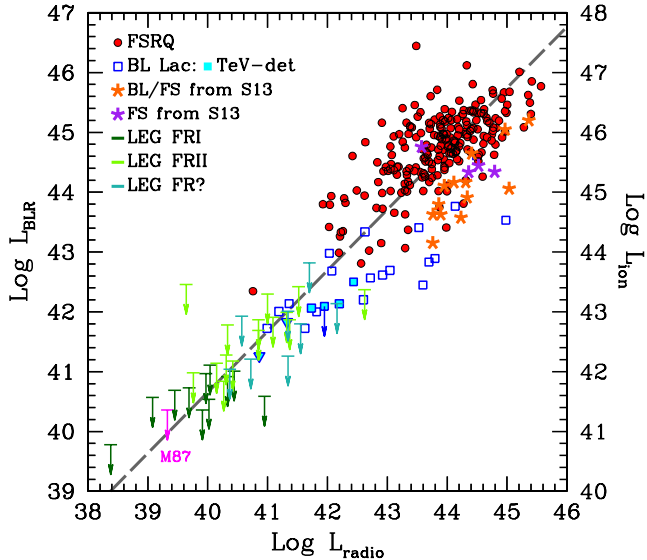,width=8.7cm,height=8.7cm}
\vskip -0.5 cm
\caption{
Broad line region luminosity as a function of radio luminosity.
Different symbols correspond to different classification of the objects. 
The dashed line is the result of a least square fit calculated among 
the detection, i.e.\ the blazars. 
}
\label{blr_radio}
\end{figure}

We aim to introduce in our study objects that do not have 
broad emission lines, but with reliable estimates of $z$ and 
$M_{\rm BH}$, and a direct proxy for the jet power. 
We also want to be able to derive an upper limit on their BLR luminosity, 
which we will use as a proxy for $L_{\rm ion}$.
As we saw from S13 BL/FS objects, the non--thermal continuum 
emitted from the jet, highly boosted because of relativistic effects, 
can dilute dramatically even strong broad emission lines. 
In the case of less luminous lines, this problem is obviously even bigger. 
In fact, the majority of $\gamma$--ray detected BL Lacs 
lacks a reliable redshift estimate, since their optical spectra are 
completely dominated by the non--thermal emission, and 
they do not show any emission features. 
This means that we cannot discriminate wether an object is genuinely lineless 
or its faint emission lines are simply not visible. 
Hence, to select only truly lineless object, we choose to introduce in 
our study a sample of radio--galaxies, i.e.\ jetted AGN in which the optical 
emission is not completely dominated by the non--thermal, boosted 
jet emission (sample description in \S\ref{sample_rg}). 
We choose a group of LEGs, to be sure that their broad 
emission lines are not present, likely because of a radiatively 
inefficient accretion disc. 
Radio--galaxies are usually not $\gamma$--ray detected, so we 
cannot use the $\gamma$--ray luminosity as a tracer of the jet power. 
We then consider radio luminosity at 8~GHz rest--frame as an alternative 
jet tracer, with the following caveat: 
the radio luminosity is emitted from the jet, and is therefore beamed 
in the emission direction. 
We will take into account the different beaming factors that characterize 
blazars and radio--galaxies in the discussion. 

\begin{figure*}
\vskip -0.6 cm 
\hskip -0.75 cm
\psfig{figure=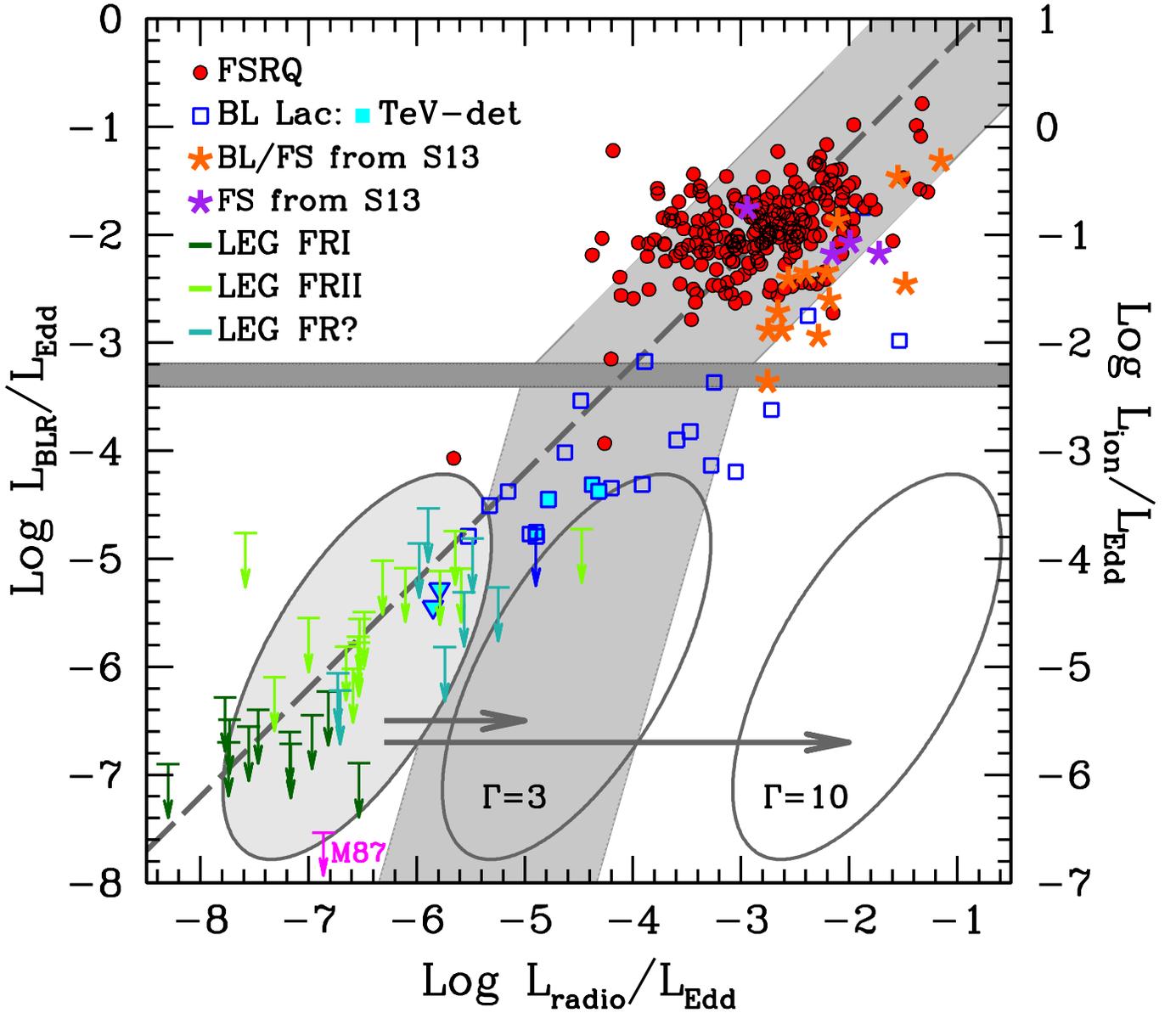,width=18.5cm,height=18.5cm}
\vskip -1 cm
\caption{Luminosity of the broad line region (in Eddington units) 
for the sources from our samples as a function of the radio 
luminosity (in Eddington units).
Different symbols correspond to different samples or a 
different classification of the sources, as labelled. 
The dashed lines indicate the bisector, rescaled to pass through the FSRQs. 
The dark grey horizontal stripe indicates the luminosity divide between 
FSRQs and BL Lacs at $L_{\rm BLR}/L_{\rm Edd}\sim5\times10^{-4}$. 
The light grey stripe indicates the expected distribution of the luminosities 
if they were produced by a Shakura--Sunyaev accretion disc for 
$L_{\rm BLR}/L_{\rm Edd}\sim5\times10^{-4}$ and an ADAF with a 
Mahadevan--like spectrum ($L_{\rm d}\propto\dot m^{3.5}$, 
see \S\ref{blr_accr} and \S\ref{discussion}). 
The leftmost ellipse include the core of the radio--galaxies of our sample. 
The central and rightmost ellipses show where the radio--galaxies 
would be located if they were beamed according to Lorentz factors 
$\Gamma=3$ or 10, respectively (as indicated by the arrows, and 
described in \S\ref{discussion}). 
}
\label{edd_radiogal}
\end{figure*}

Figure \ref{blr_radio} shows the comparison between broad line region 
and radio luminosities in all the sources of our samples. 
All the radio--galaxies have upper limits on their $L_{\rm BLR}$ values, 
since they are explicitly selected to be lineless (see \S\ref{blr} for the 
upper limits derivation). 
Note that the radio luminosities calculated for blazars and radio--galaxies 
(and plotted in Figure \ref{blr_radio}) are differently beamed, 
because of different viewing angles. 
Therefore, the linear correlation over the whole luminosity range is 
only apparent. 
To properly study the $L_{\rm radio}-L_{\rm BLR}$ relation, we have 
to homogenize the beaming factors. 
This is true also if we consider the two luminosities normalized to the 
Eddington luminosity, as shown in Figure \ref{edd_radiogal}, 
which we analyze in detail in \S\ref{discussion}. 

Note that 
the objects reclassified as BL/FS are located at the highest 
radio luminosity edge of the correlation in both Figure \ref{blr_radio} and \ref{edd_radiogal}. 
This clearly highlights their FSRQ nature, associated with an 
uncommonly luminous synchrotron emission, very well traced by 
the radio luminosity itself. 
They can easily be considered as the tail at high magnetic field 
energy density of the class of FSRQs. 

Note that in both plots the radio--galaxies are located at $L_{\rm BLR}$
(normalised by $L_{\rm Edd}$ and not) lower than the BL Lacs, 
with a small overlap.
By including these objects, we finally manage to achieve very low disc luminosities, 
and hence very low accretion rates, which is crucial for our understanding. 
We discuss the implications of this in \S\ref{discussion}.
The upper limits hint also at a change in relationship between the tracers of $L_{\rm ion}$ and jet power
 but their presence in the low--luminosity branch 
of the sample prevents us from a proper parametric characterization of this relation. 
To understand if the relationship between $L_{\rm BLR}/L_{\rm Edd}$ and $L_{\rm radio}/L_{\rm Edd}$ 
is better explained by a single or a broken power--law, we do the following: 
we treat the upper limits as detections and, since these cluster mostly at the low power end,
 the slope of the power--law we get from the fit has to be considered as a lower limit (i.e., the true
 value will be steeper). With this approach, we can compare the two hypotheses 
of a single and a broken power--law by using an F--test. 
By minimizing the $\chi^2$ values in the hypotheses of a single power--law 
($\log L_{\rm BLR}/L_{\rm Edd}=(0.87\pm0.12)\log L_{\rm radio}/L_{\rm Edd}+0.28$)
and a broken power--law with break at $L_{\rm radio}/L_{\rm Edd}\sim-4$, 
we infer that the data are better described by a broken power--law 
with a $99.97\%$ level of confidence\footnote{
The uncertainties on $\log L_{\rm BLR}/L_{\rm Edd}$, necessaries to calculate the $\chi^2$, 
are derived from the uncertainties on the broad line luminosities and on the 
black hole mass measurements, and are typically $\sim$0.3dex. }.
The break value is fixed to correspond to the dividing value between FSRQs and BL Lacs 
in $L_{\rm BLR}/L_{\rm Edd}$ discussed in \S\ref{gamma}, following the single power--law.  
The broken power--law that better fits our data is described by the relation:
\begin{equation}
\log\frac{L_{\rm BLR}}{L_{\rm Edd}} = \left\{
\begin{array}{ll}
(0.98\pm0.02) \log\frac{L_{\rm radio}}{L_{\rm Edd}} +0.72; &  \log\frac{L_{\rm radio}}{L_{\rm Edd}}<-4 \\
(0.81\pm0.03) \log\frac{L_{\rm radio}}{L_{\rm Edd}} +0.04; &  \log\frac{L_{\rm radio}}{L_{\rm Edd}}\geq -4 
\end{array} \right.
\end{equation}
This shows that a single power law is not a good representation of our data. 
We discuss in the next section the meaning of this result.

\section{Discussion}
\label{discussion}

Figure \ref{edd_radiogal} is the main result of our work. 
As we have already pointed out, the radio luminosity has a  different physical meaning 
in the case of blazars and radio--galaxies, because of different beaming levels. 
To properly compare them, we have to beam the radio luminosity of the radio--galaxies, 
assuming an average bulk Lorentz factor $\Gamma$ and a viewing angle $\theta$. 
This will shift them at higher radio luminosities, rejoining them with their aligned analogous AGN. 
Note that the different orientation of blazars and their parent population does not affect the 
BLR luminosity, since it is emitted isotropically. 
Therefore, the parent population of a group of blazars would be located at the same 
$L_{\rm BLR}/L_{\rm Edd}$, with a $L_{\rm radio}/L_{\rm Edd}$ smaller than the 
corresponding aligned blazars. 
From their position in Figures \ref{blr_radio} and \ref{edd_radiogal}, the LEG FRI radio--galaxies 
are not the parent population of the BL Lacs included in our study (and see Chiaberge et al.\ 2000), 
In fact, it is important to remember that the BL Lacs in our sample 
have broad emission lines, while the FRI we collected are intrinsically without broad emission lines. 
This spectral difference explains why our FRI and our BL Lacs are intrinsically different. 
There is likely a population of {\it truly lineless BL Lacs} of which these LEGs 
are actually the parent population, that we are not able to include in our study.  
The only upper limits on $L_{\rm BLR}$ that we derived for a group of BL Lacs in TS12 
(from Plotkin et al.\ 2011) were anyway located in the same $L_{\rm BLR}$ range as the 
broad--line BL Lacs. 
None of the known BL Lacs with a measured redshift represent the re--oriented analogous 
of the LEG FRI. 

However, we are considering only the tip of the iceberg of the BL Lac population: 
2LAC includes 475 BL Lacs, and most of them do not have a reliable redshift estimate, 
since their optical spectra are completely featureless. 
Without a redshift estimate, we cannot derive their intrinsic luminosity in any band, 
nor calculate an upper limit on their broad line luminosity. 
Therefore, they cannot be compared to the other blazars in our work. 
Among them, there are most likely the truly lineless BL Lacs that would be 
necessary to study the very low accretion regimes, and of course they 
would be the aligned analogous to the LEGs that we include in our study. 
This makes the radio--galaxies without broad emission lines even more relevant for our work, 
since they are the only valid tracer of the low--accretion regime. 
But to use them to explore that regime, we have to uniform their beaming to the blazar one. 

How much do we have to beam the radio luminosity of the LEGs in our sample 
to compare them with blazars? 
The beaming factor of a source with generic $\Gamma$, $\beta$ and $\theta$ is:
\begin{equation}
\delta = \frac{1}{\Gamma (1-\beta\cos\theta)}
\end{equation}
We take an average viewing angle $\theta_{\rm LEG}\sim40^\circ$, assuming that the LEGs 
are misaligned. 
We want to beam their radio luminosity as they were oriented as a blazar, 
i.e.\ with $\theta_{\rm BL}\sim3^\circ$. 
Therefore we have to boost the radio luminosity by a factor:
\begin{equation}
\left( \frac{\delta_{\rm BL}}{\delta_{\rm LEG}} \right)^3 = 
\left( \frac{1-\beta\cos\theta_{\rm LEG}}{1-\beta\cos\theta_{\rm BL}} \right)^3
\end{equation}
If we assume their jets are beamed with a Lorentz factor similar to a common blazar, i.e.\ 
$\Gamma\simeq10$, this beaming factor would be $\sim5\times10^4$, shifting the 
LEGs as drawn in Figure \ref{edd_radiogal} (rightmost ellipse). 
None of the existing blazars could populate that region of the plot, even considering the 
BL Lacs without redshift. 
In fact, if we assign an arbitrary $z=1$ to all BL Lacs without a redshift 
estimate in 2LAC, we obtain an average radio luminosity at 8~GHz rest--frame 
$\log L_{\rm radio}\sim43.2$, that would correspond to 
$\log(L_{\rm radio}/L_{\rm Edd})\sim-3.3$. 
This would not be enough to populate the region occupied by the rightmost ellipse, 
i.e.\ the radio--galaxies beamed of a factor $\sim5\times10^4$ due to $\Gamma\simeq10$. 
To populate that region, one should postulate the existence of AGN with an extremely 
powerful and relativistic jet, associated with an accretion structure extremely radiatively inefficient, 
or even without any accretion structure present. 
This beaming level seems then quite unlikely. 

We can consider another beaming option. 
From VLBI studies, there is evidence that in strong TeV BL Lacs the pc--scale jets 
move slowly (Edwards \& Piner 2002; Piner \& Edwards 2004). 
At the same time, the extreme variability of their intense TeV luminosity 
implies that the jet should be highly relativistic, at least in the region where the TeV 
emission originates. 
To justify such a discrepancy, the two less demanding hypothesis that have been 
advanced are: (i) a deceleration of the emitting region between the TeV and the radio 
locii (Georganopoulos \& Kazanas 2004); and (ii) a spine--layer structure 
of the jet (Ghisellini et al.\ 2005). 
Furthermore, detailed observations performed with the VLBI show a morphology that 
suggests the presence of a slower external layer, surrounding a faster core 
in the jet in the lineless BL Lac Mkn 501 (Giroletti et al.\ 2004). 
Similar results have also been obtained in the case of some radio--galaxies 
(Giovannini et al.\ 1999; Swain et al.\ 1998; Owen et al.\ 1989).
Moreover, a velocity structure helps in explaining other features typical 
of radio--galaxies, such as the configuration of their magnetic field 
(Komissarov 1990; Laing 1993). 
According to this hypothesis, the radio emission should then be characterized 
by a rather small Lorentz factor $\Gamma\sim3$, being emitted by the 
external layer. 
In this case, the LEG radio luminosity can be boosted by a smaller factor  
$(\delta_{\rm BL}/\delta_{\rm LEG})^3\sim100$. 
 
The central ellipse in Figure \ref{edd_radiogal} represents this hypothesis. 
Such a beaming factor shifts the radio power of the LEGs by such an amount that 
in the $L_{\rm BLR}/L_{\rm Edd}$ -- 
$L_{\rm radio}/L_{\rm Edd}$ plane they now follow a much steeper relationship than the 
almost linear best fit derived in \S\ref{radio} with the broken power--law. 
As previously done, we perform an F--test to compare the single and 
a broken power--law hypotheses, obtained as best fits of our data, all assumed as detections. 
The two best fits are 
$\log L_{\rm BLR}/L_{\rm Edd}=(0.89\pm0.19)\log L_{\rm radio}/L_{\rm Edd}+0.36$
and again a broken power--law with break at $L_{\rm radio}/L_{\rm Edd}\sim-4$:  
\begin{equation}
\log\frac{L_{\rm BLR}}{L_{\rm Edd}} = \left\{
\begin{array}{ll}
(2.00\pm0.06) \log\frac{L_{\rm radio}}{L_{\rm Edd}} +4.8; &  \log\frac{L_{\rm radio}}{L_{\rm Edd}}<-4 \\
(0.78\pm0.03) \log\frac{L_{\rm radio}}{L_{\rm Edd}} -0.08; &  \log\frac{L_{\rm radio}}{L_{\rm Edd}}\geq -4 
\end{array} \right.
\end{equation}
The F--test shows that, again, a broken power--law provides a better description of the data
at the $>99.99\%$ level. 
We stress again that the slope derived with this method at values below the break provides 
{\it only a lower limit} to the true slope. 
In other words, at luminosities smaller than the break, jetted AGN follow a relation steeper than: 
\begin{equation}
\frac{L_{\rm BLR}}{L_{\rm Edd}} \propto \left( \frac{L_{\rm radio}}{L_{\rm Edd}} \right)^2
\end{equation}

This clearly means that the jetted AGN highlight the break that we expect from standard 
theory of accretion discs, i.e.\ there is a change in the accretion process at 
$L_{\rm BLR}/L_{\rm Edd}\approx10^{-3.5}$ (the light grey stripe in Figure \ref{edd_radiogal}). 
In fact, if we interpret the LEGs as the jetted AGN with lowest accretion rate, as suggested 
by their $L_{\rm BLR}$, they would follow a relation that is steeper than the one expected in 
the case of an efficient accretion structure (i.e.\ linear).
This relation is even steeper than the simple dependence of $L_{\rm d}$ on the accretion rate 
($L_{\rm d}\propto\dot M^2$), since the slope we derive is only a lower limit on the true slope. 
This result is consistent with highly radiatively inefficient models, closer to the most inefficient 
one, i.e.\ a pure ADAF, from which we expect the relation  
\begin{equation}
\frac{L_{\rm BLR}}{L_{\rm Edd}} \propto \frac{L_{\rm ion}}{L_{\rm Edd}} 
\propto \dot m^{3.5}. 
\end{equation} 
below the value of $\dot m_c$, which separates the two accretion regimes. 

We find that such a transition occurs 
at $\dot m_c\sim0.1$, i.e.\ $L_{\rm BLR}/L_{\rm Edd}\sim 5\times10^{-4}-10^{-3}$ 
if a radiative efficiency $\eta\sim0.1$ is assumed. 
This threshold is consistent also with the accretion rate transition 
between FRI and FRII found by Ghisellini \& Celotti (2001). 
The hypothesis of a transition at $\dot m_c\sim10^{-4}$ would not be consistent with 
the beamed LEG data. 
In any case, we do not expect a sharp transition, but more likely a smooth one, since we do not 
observe a clear bimodality in the $L_{\rm BLR}/L_{\rm Edd}$ distribution.

\section{Conclusions}
\label{conclusion}

In this work, we have explored the connection between jet and 
accretion structure in jetted AGN, using 267 broad emission line blazars 
and 38 broad--line--less radio--galaxies, all with known redshift, a measure 
of the jet power and an estimate of the black hole mass. 
In the case of blazars, we have used both $\gamma$--ray and radio luminosities 
to trace their jet power, while the radio--galaxies only have the radio 
core power as a jet tracer. 
Since they do not show broad emission lines, we have derived robust upper 
limits on their broad line region luminosity from the luminosity of their narrow lines. 
They are crucial to explore the low--accretion regime of jetted AGN. 
The results we obtained can be summarized as follows:

\begin{enumerate}

%
\item
With a sample composed by both blazars and radio--galaxies, we 
finally can identify the transition between efficient and inefficient 
accretion structures. 
With only blazars, we are not able to include the very low--accreting 
objects, since they would be lineless and dominated by the jet 
non--thermal emission, and therefore again without a redshift estimate. 
LEG radio--galaxies are therefore the only mean to study the 
radiatively inefficient accretion regime. 

\item
The most reasonable beaming option for the radio--galaxies we included is due to jets 
structured with a central extremely relativistic spine, surrounded by a slower layer.
A high Lorentz factor $\Gamma=10$, necessary to justify some observational 
properties, would characterize only the central part of the jet. 
A slower layer likely surrounds this extreme spine, and would be the 
responsible of the radio emission from the jet. 
This external layer is characterized 
by a smaller Lorentz factor  ($\Gamma\sim3$), implying a smaller 
beaming factor to homogenize radio--galaxies to blazars.  

\item
The transition between efficient and inefficient accretion regimes 
occurs at the standard critical value $\dot m_c\sim0.1$, 
i.e.\ at $L_{\rm BLR}/L_{\rm Edd}\sim5\times10^{-4}-10^{-3}$ 
assuming an accretion efficiency $\eta\sim0.1$. 
At accretion values lower than that, the ionizing luminosity decreases  
with a slope steeper than $\propto\dot m^{2}$, clearly traced by the radio--galaxies. 
This is consistent with a transition from an efficient to an inefficient regime at low accretion rates. 
A relevant decrease in the ionizing luminosity is in fact expected in all the highly inefficient 
accretion regimes (e.g.\ the ADAF model).

\end{enumerate}

\section*{Acknowledgments}
We thank the referee for useful comments that improved the paper.
Part of this work is based on
archival data, software or online services provided by the ASI Data
Center (ASDC).
This research made also use of the NASA/IPAC Extragalactic Database (NED) 
which is operated by the Jet Propulsion Laboratory, Caltech, under contract 
with NASA.



\section*{Appendix}


\begin{figure*}
\centering
\hskip -0.2 cm
\psfig{figure=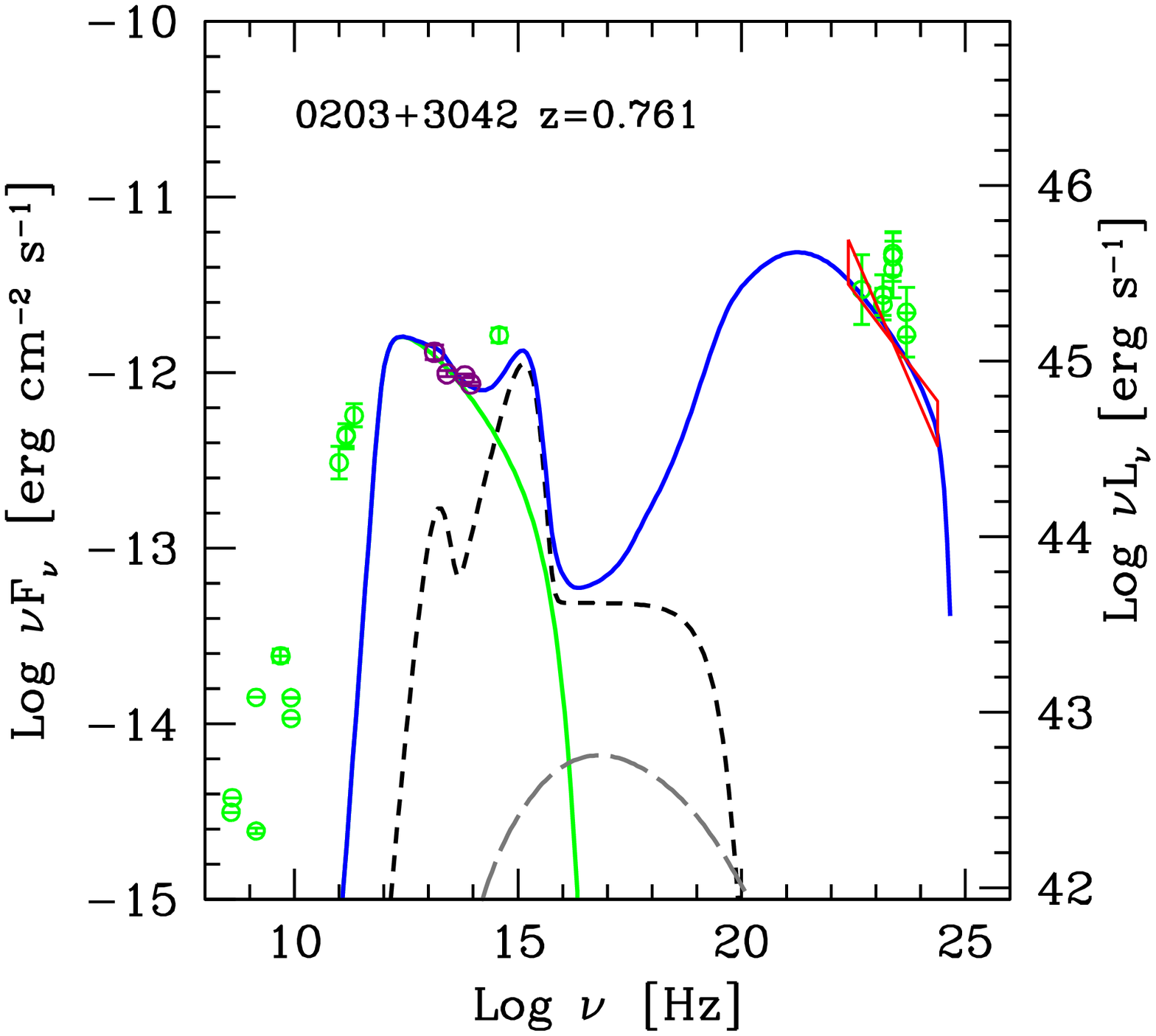,width=6cm,height=5.8cm}
\hskip -0.2 cm
\psfig{figure=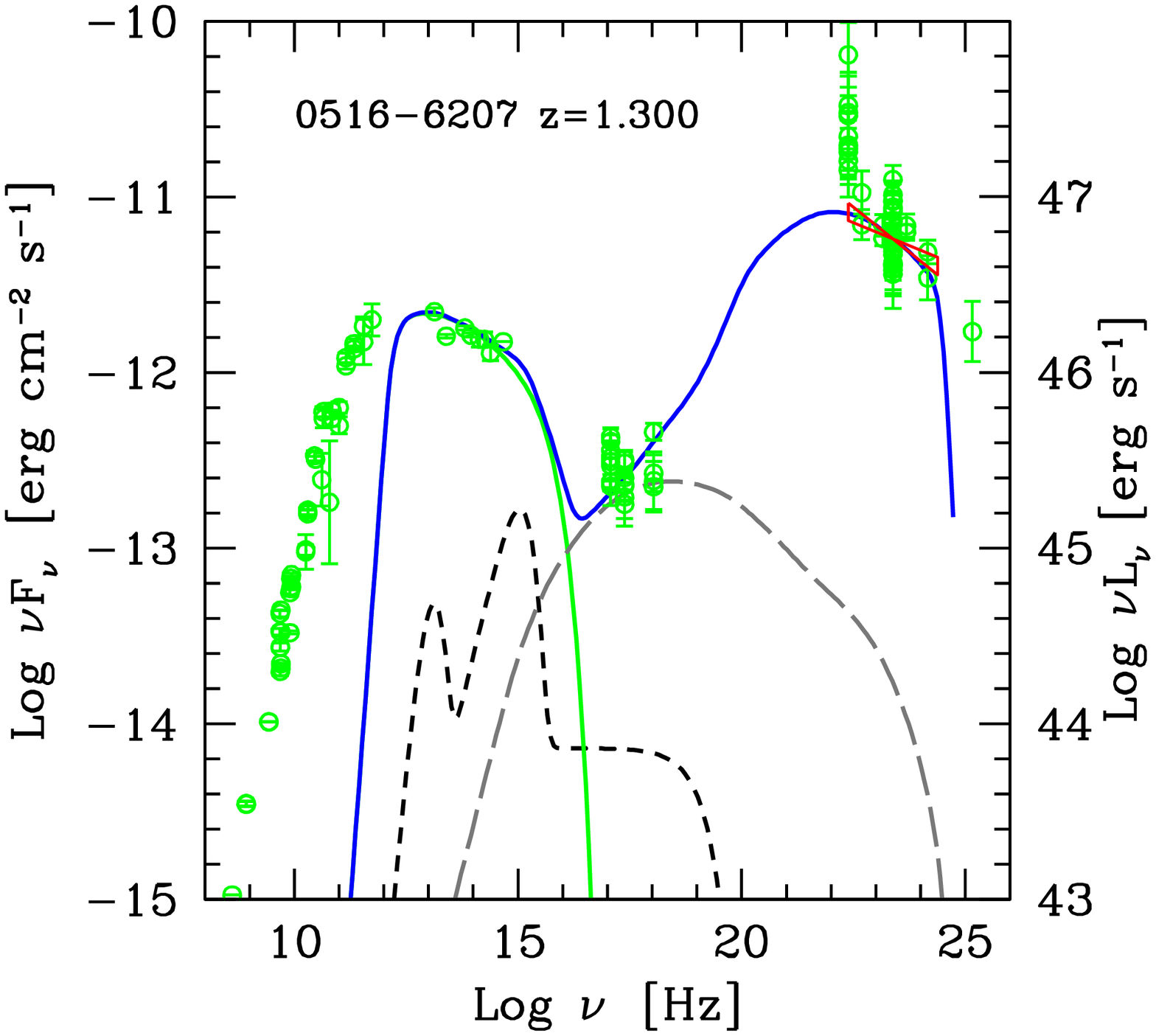,width=6cm,height=5.8cm}
\vskip -0.6 cm
\hskip -0.2 cm
\psfig{figure=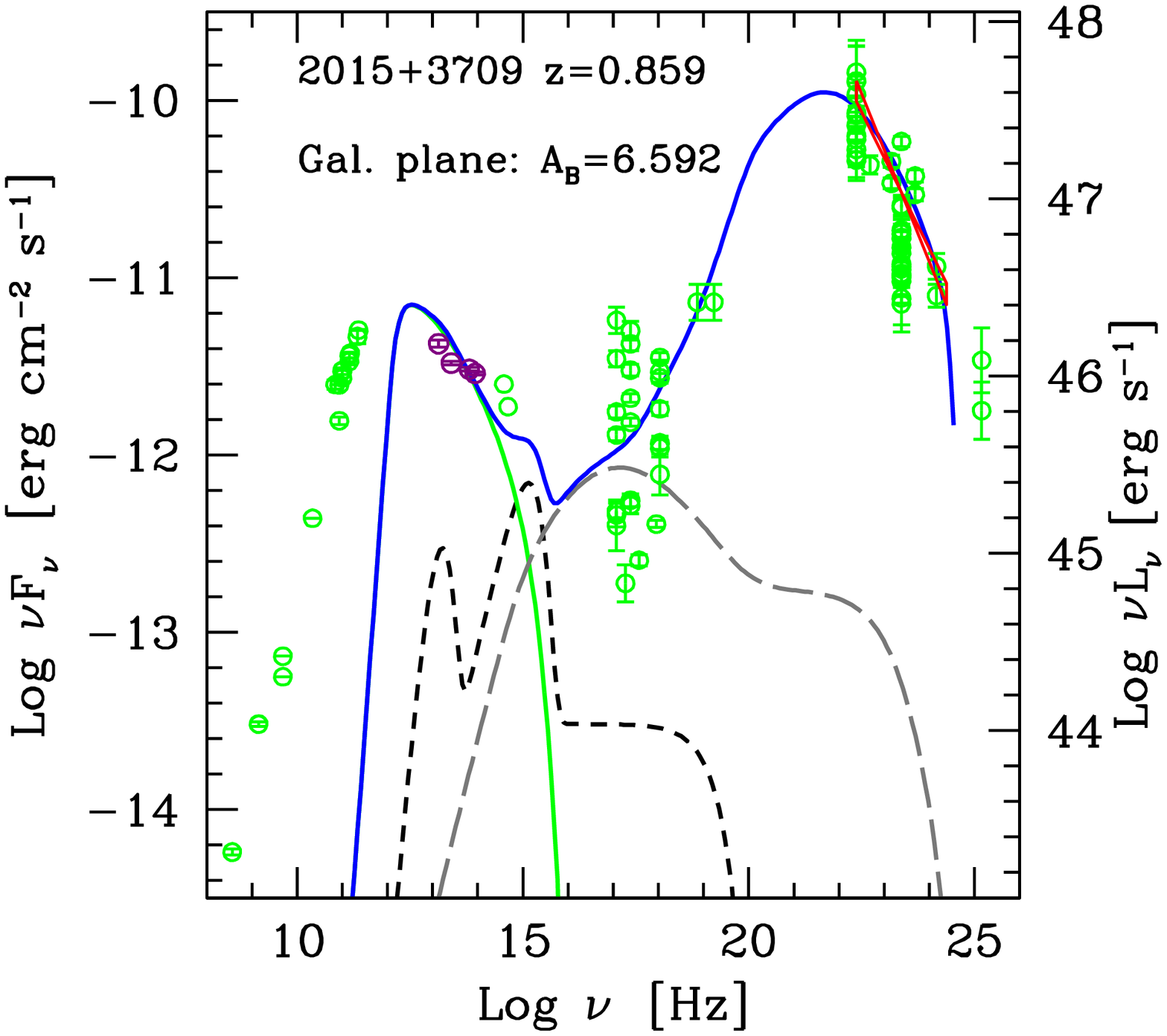,width=6cm,height=5.8cm}
\hskip -0.2 cm
\psfig{figure=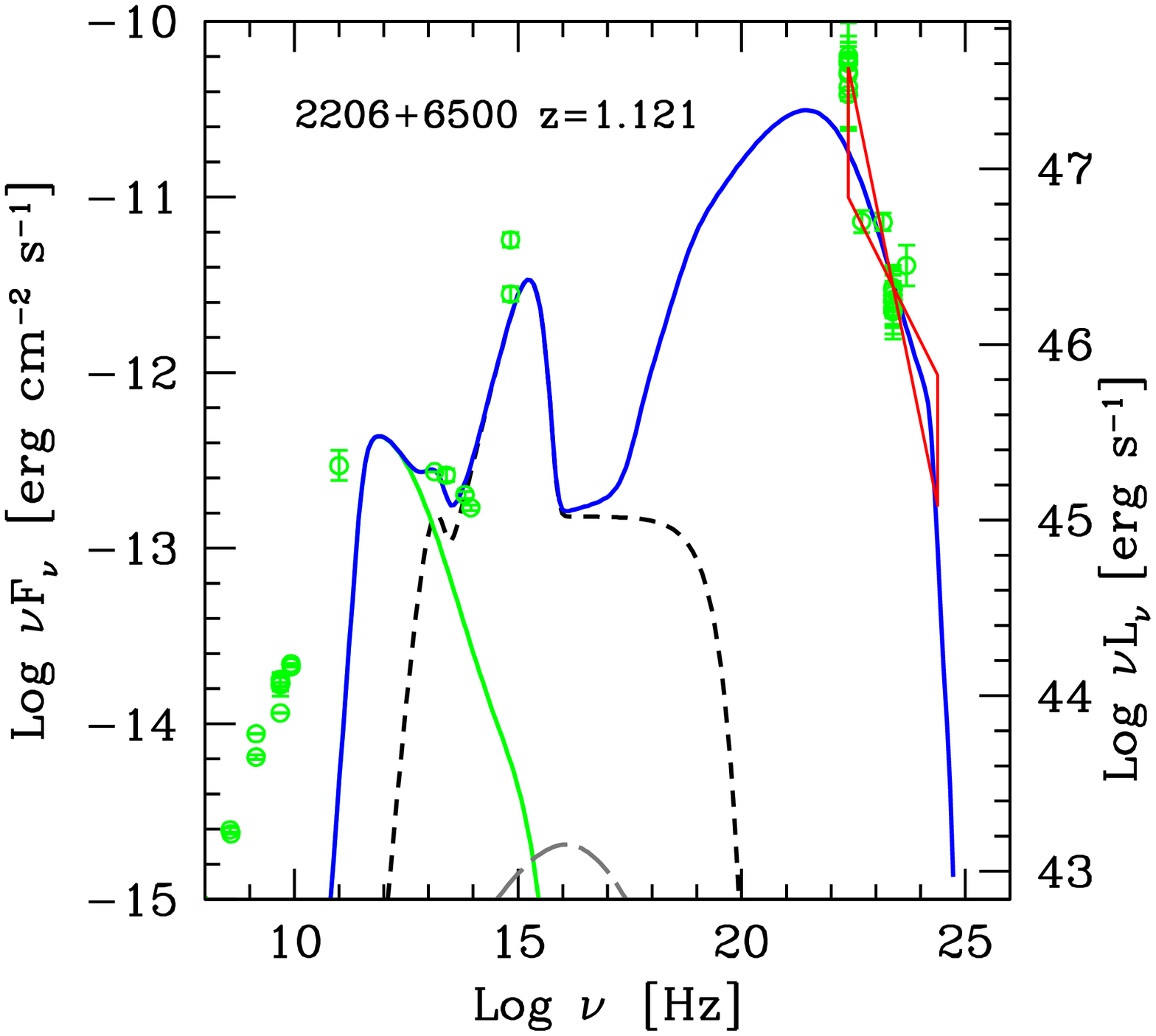,width=6cm,height=5.8cm}
\vskip -0.5 cm
\caption{Sources reclassified as FS. 
The blue solid lines represent the overall models. 
The green solid line is the synchrotron components, the black dashed 
line is the thermal emission from accretion disc, torus and corona, 
while the grey long--dashed line is the Synchrotron--Self--Compton emission. 
}
\label{sed_fs}
\end{figure*}

\begin{figure*}
\centering
\hskip -0.2 cm
\psfig{figure=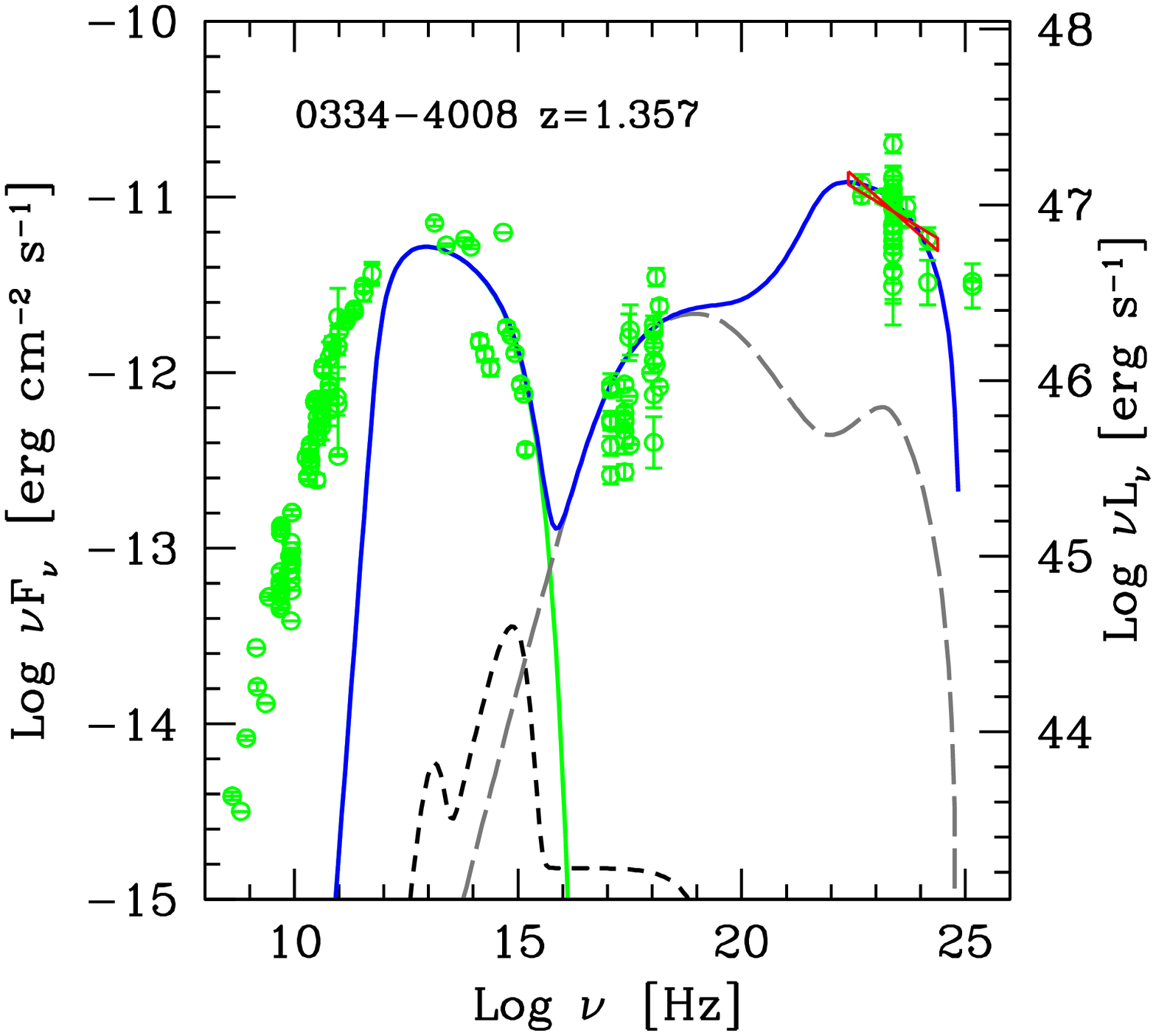,width=6cm,height=5.8cm}
\hskip -0.2 cm
\psfig{figure=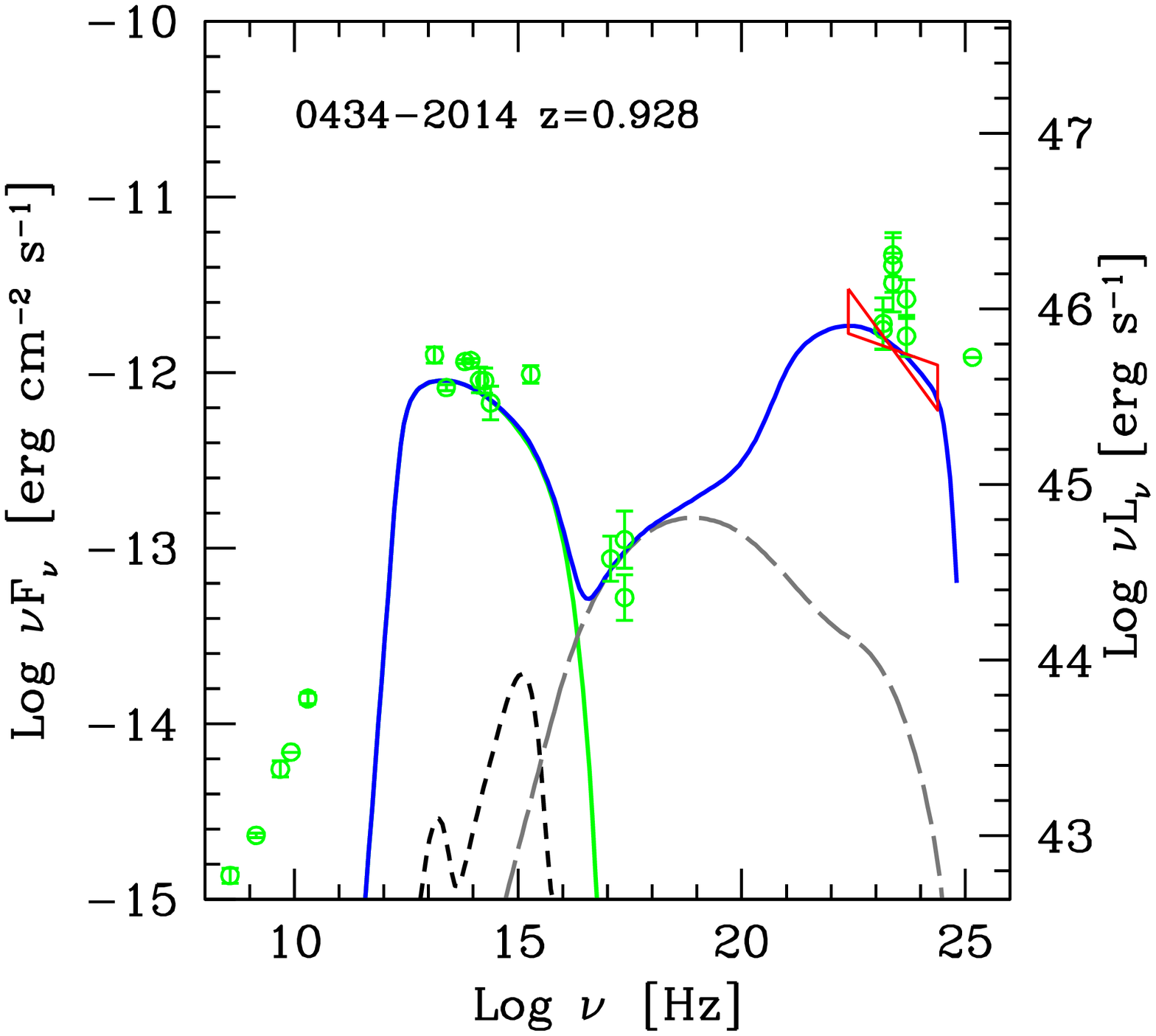,width=6cm,height=5.8cm}
\hskip -0.2 cm
\psfig{figure=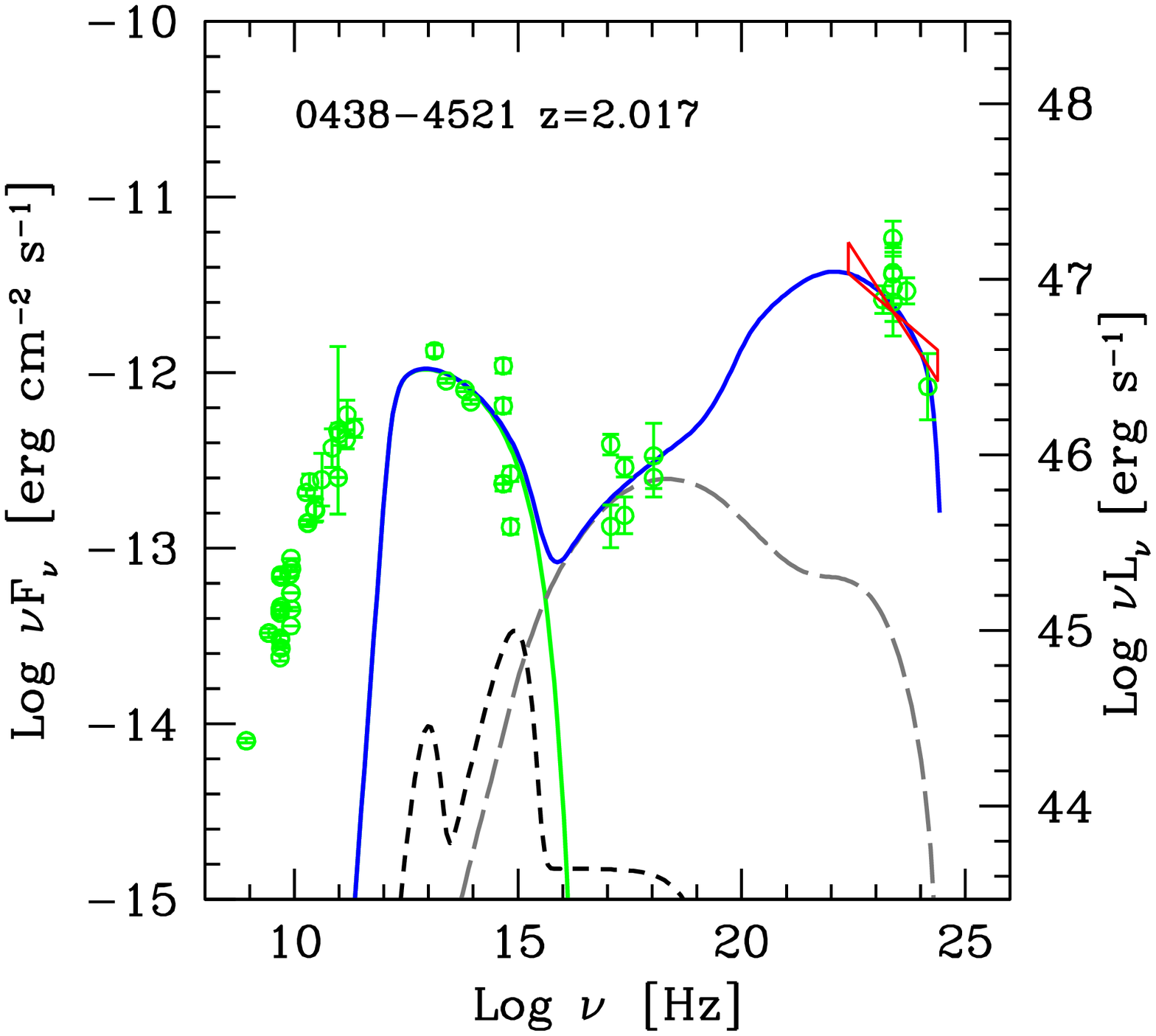,width=6cm,height=5.8cm}
\vskip -0.6 cm
\hskip -0.2 cm
\psfig{figure=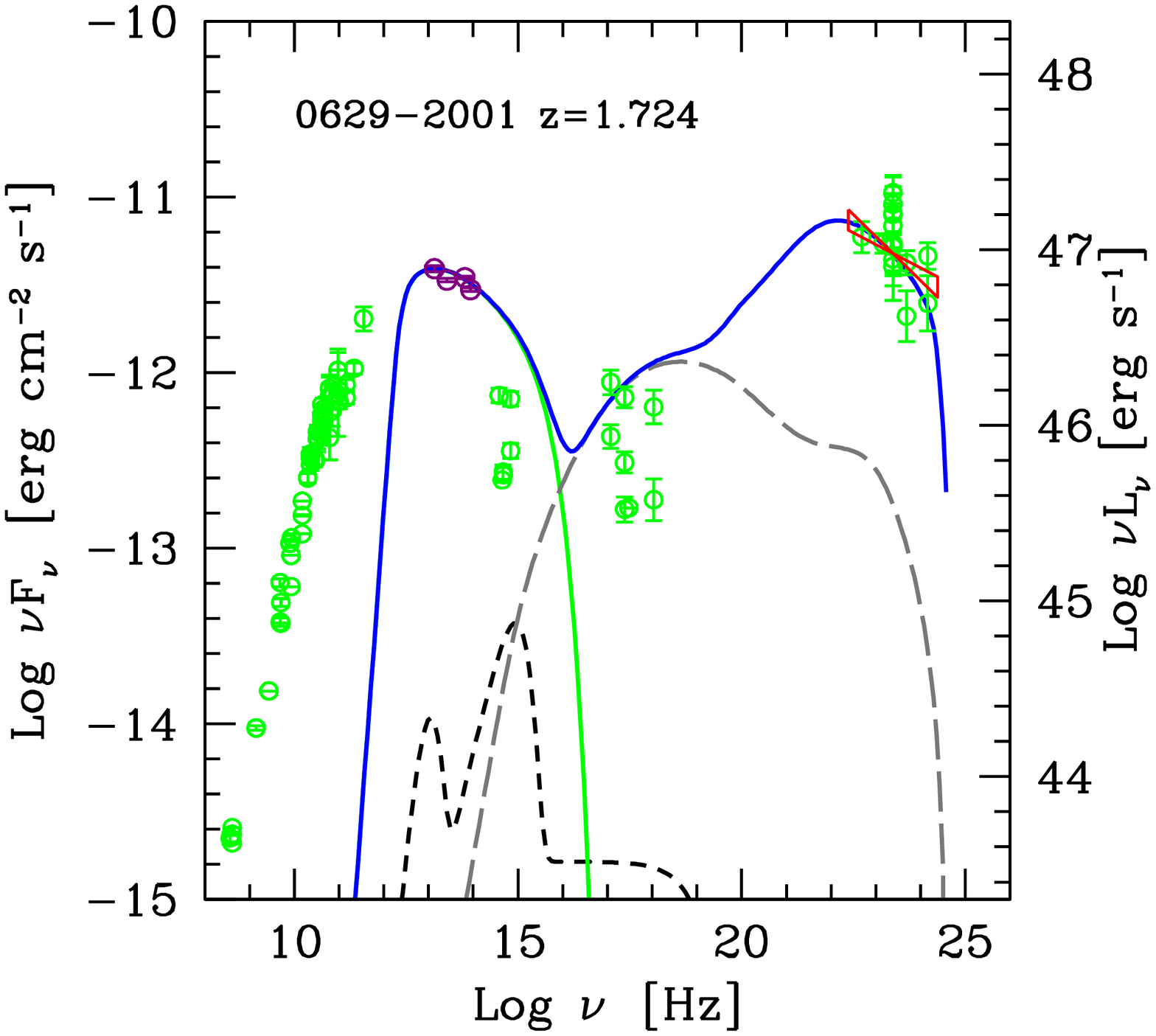,width=6cm,height=5.8cm}
\hskip -0.2 cm
\psfig{figure=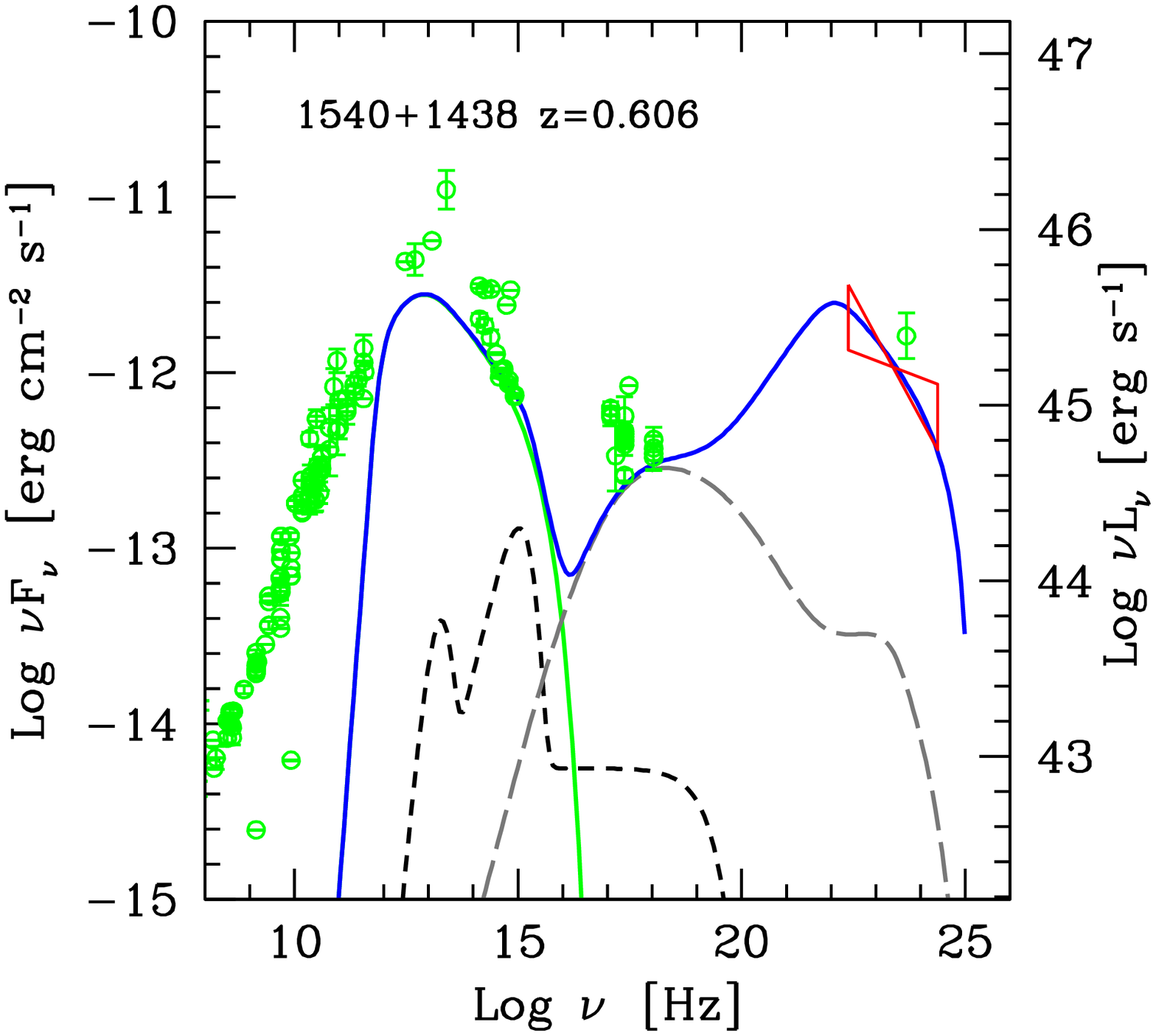,width=6cm,height=5.8cm}
\hskip -0.2 cm
\psfig{figure=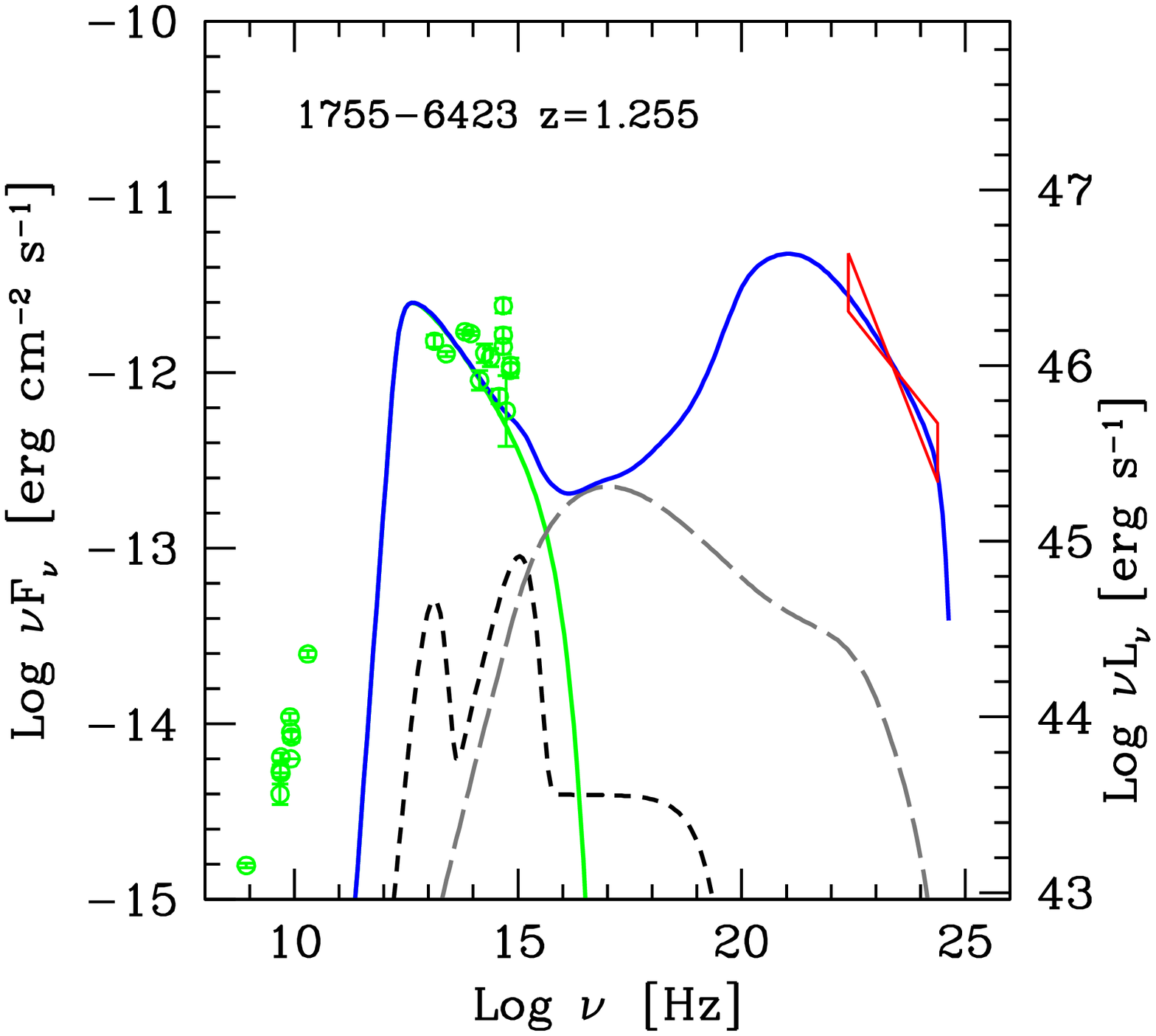,width=6cm,height=5.8cm}
\vskip -0.6 cm 
\hskip -0.2 cm
\psfig{figure=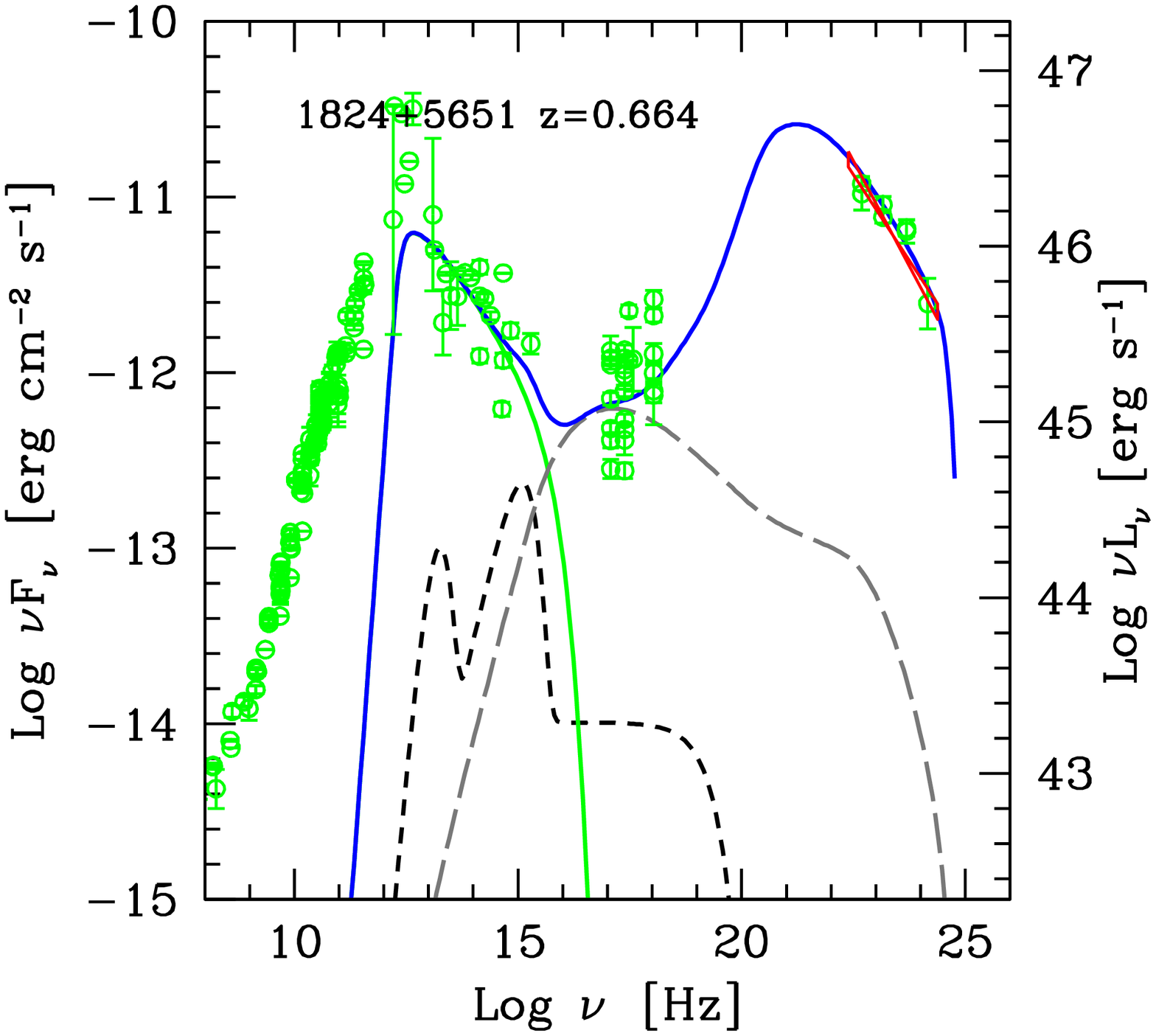,width=6cm,height=5.8cm}
\hskip -0.2 cm
\psfig{figure=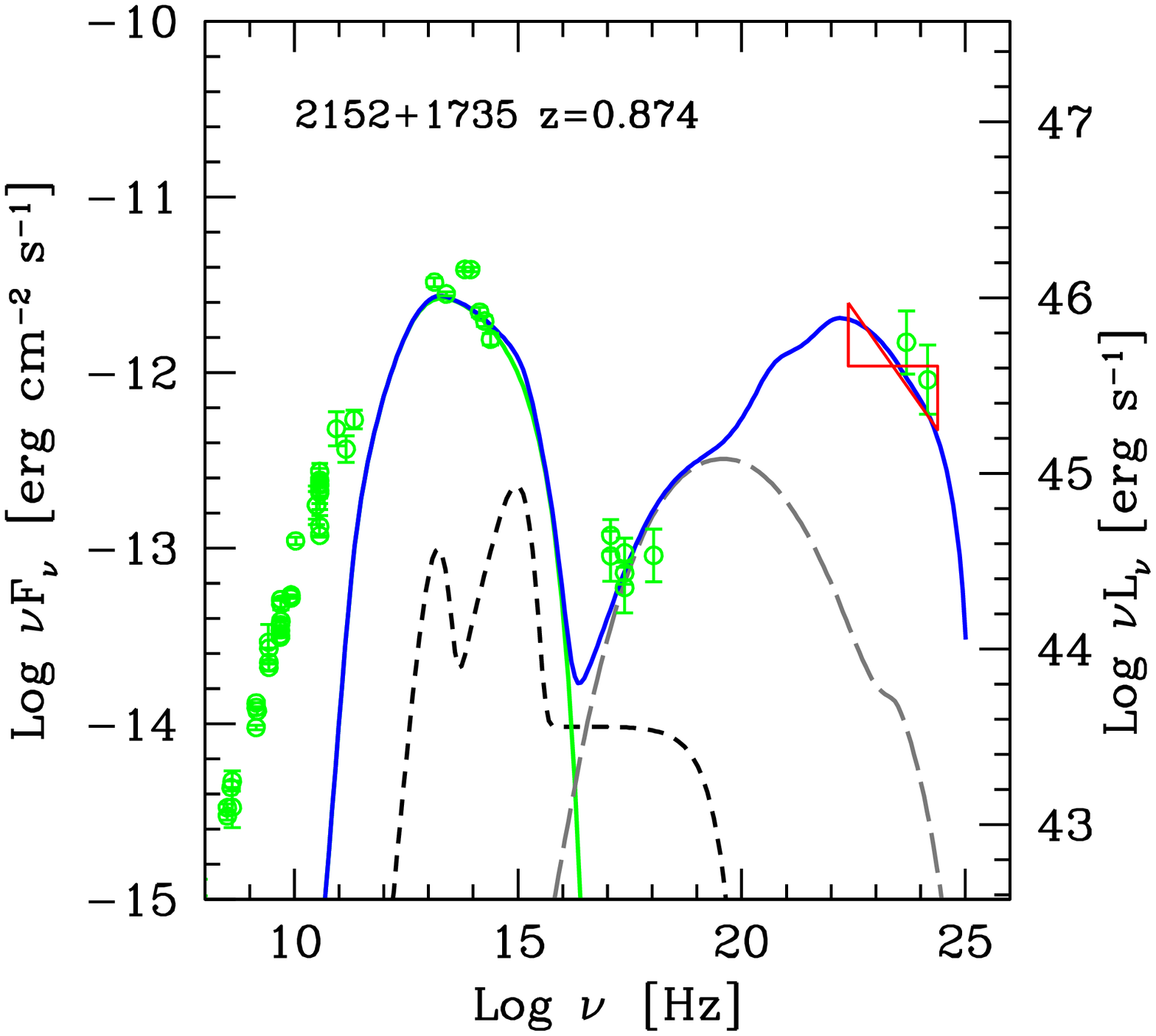,width=6cm,height=5.8cm}
\hskip -0.2 cm
\psfig{figure=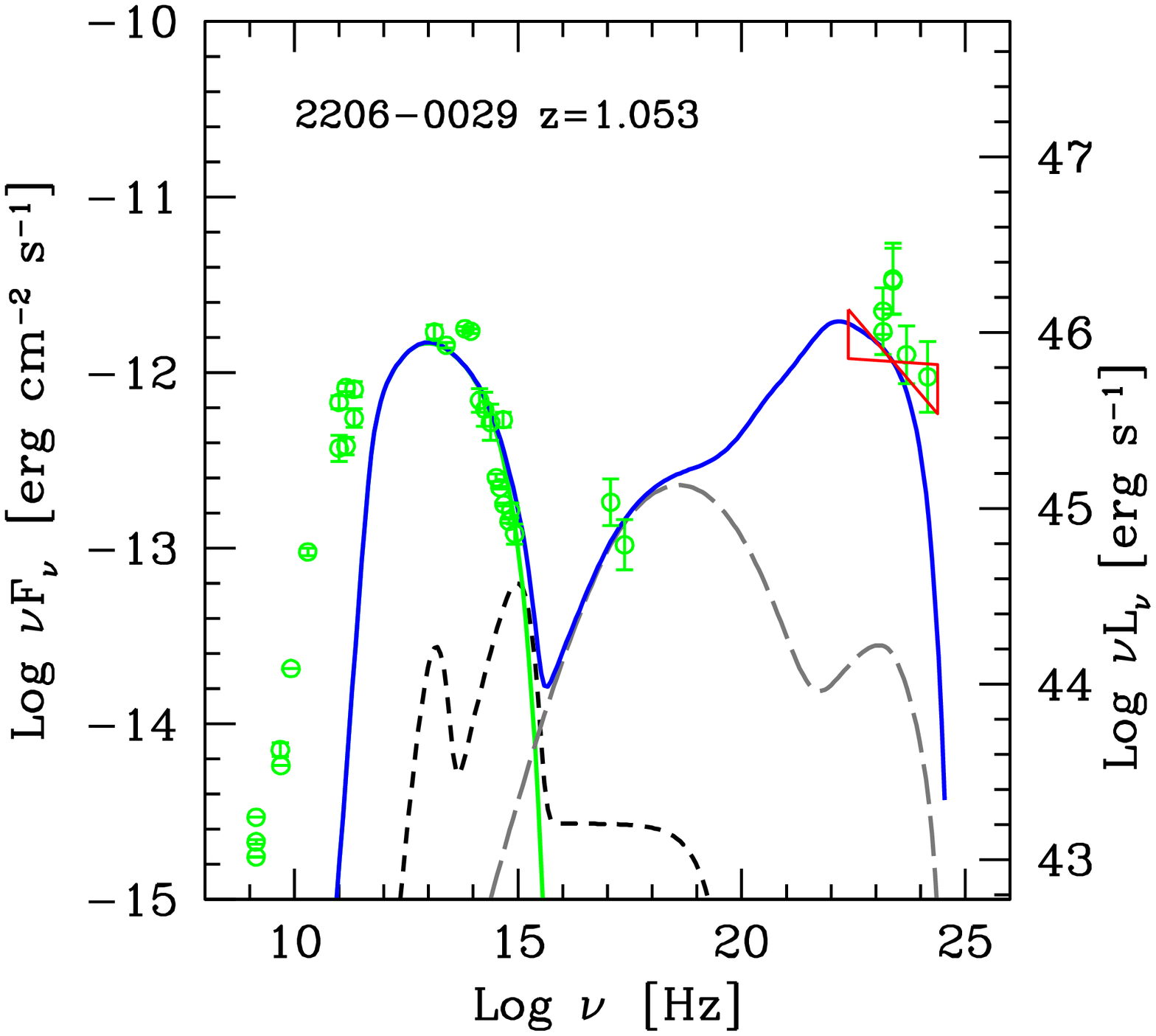,width=6cm,height=5.8cm}
\vskip -0.6 cm 
\hskip -0.2 cm
\psfig{figure=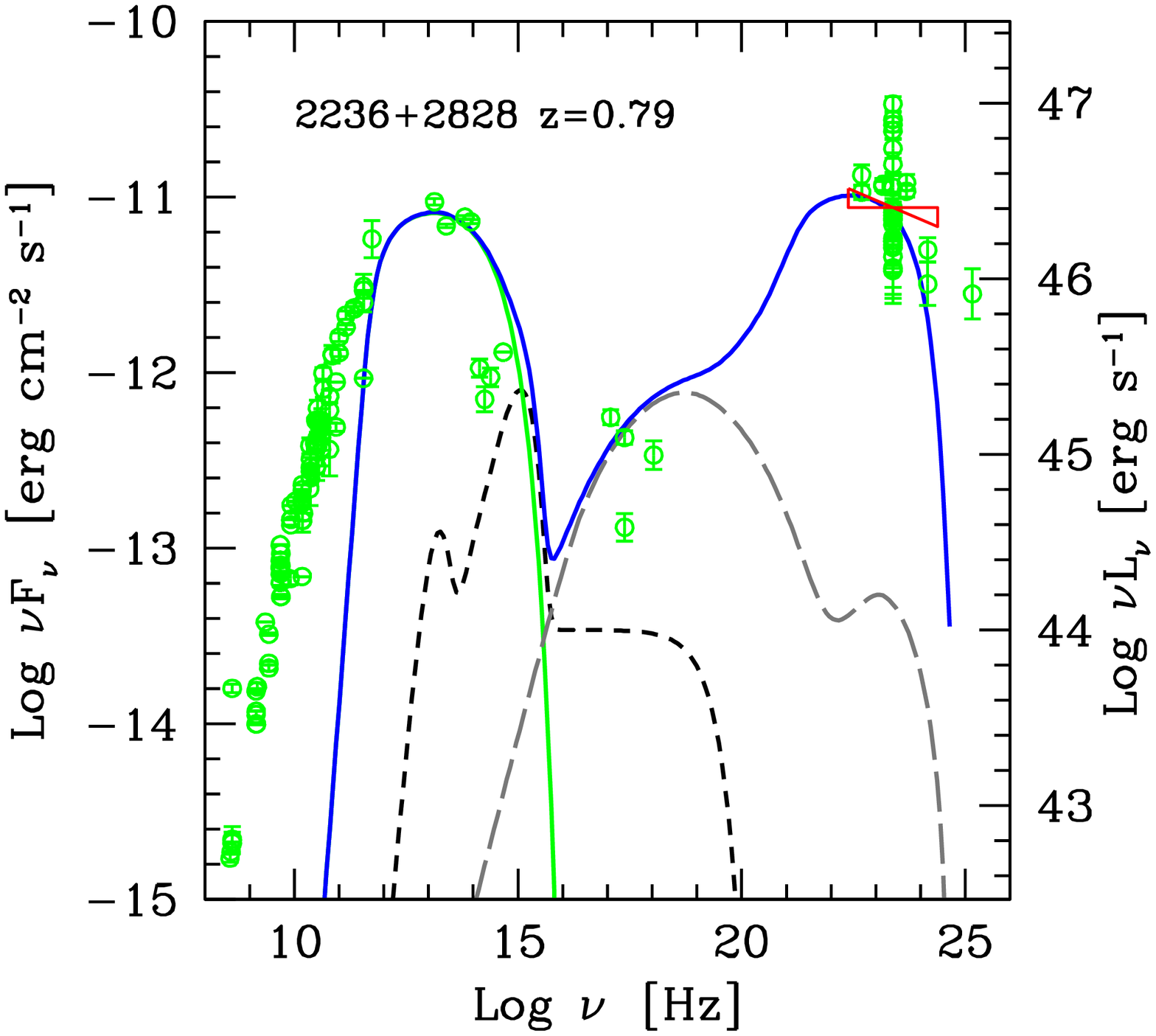,width=6cm,height=5.8cm}
\vskip -0.5 cm
\caption{Sources reclassified as BL/FS, with optical spectra showing 
weak broad emission lines, but SEDs more typical of FSRQs. 
Lines as in Figure \ref{sed_fs}.
}
\label{sed_blfs}
\end{figure*}

\begin{figure*}
\centering
\vskip -0.6 cm 
\hskip -0.2 cm
\psfig{figure=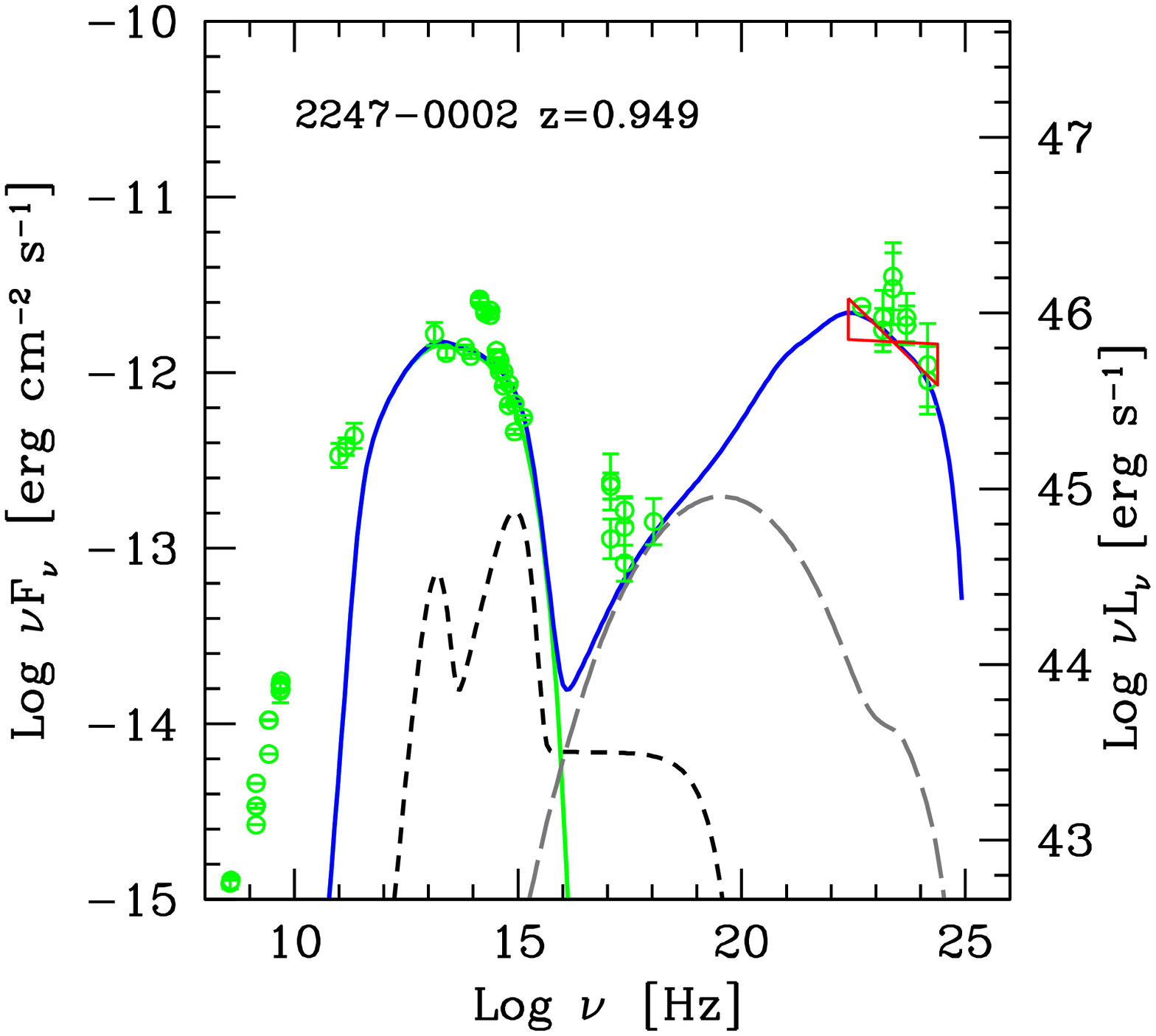,width=6cm,height=5.8cm}
\hskip -0.2 cm
\psfig{figure=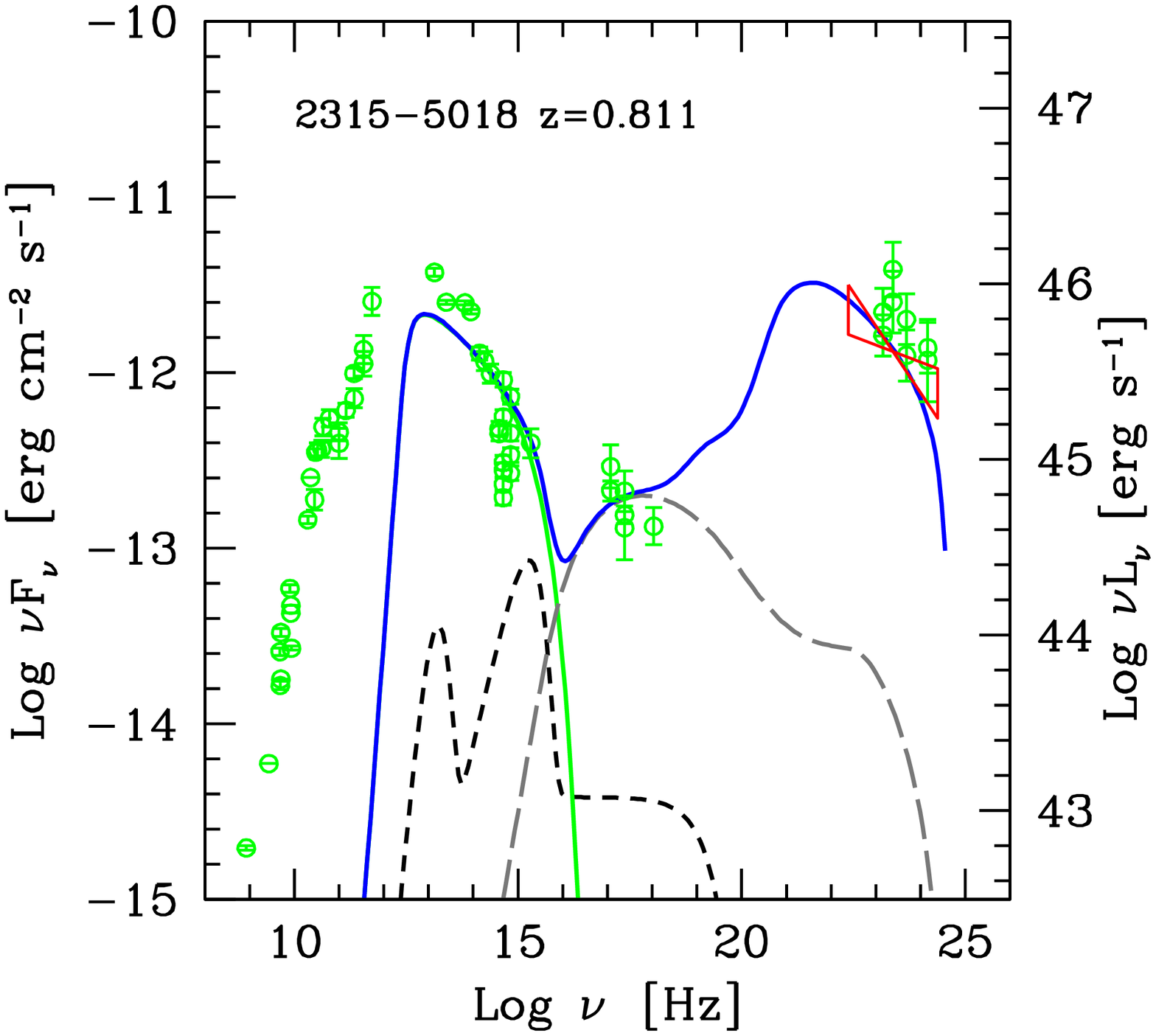,width=6cm,height=5.8cm}
\hskip -0.2 cm
\psfig{figure=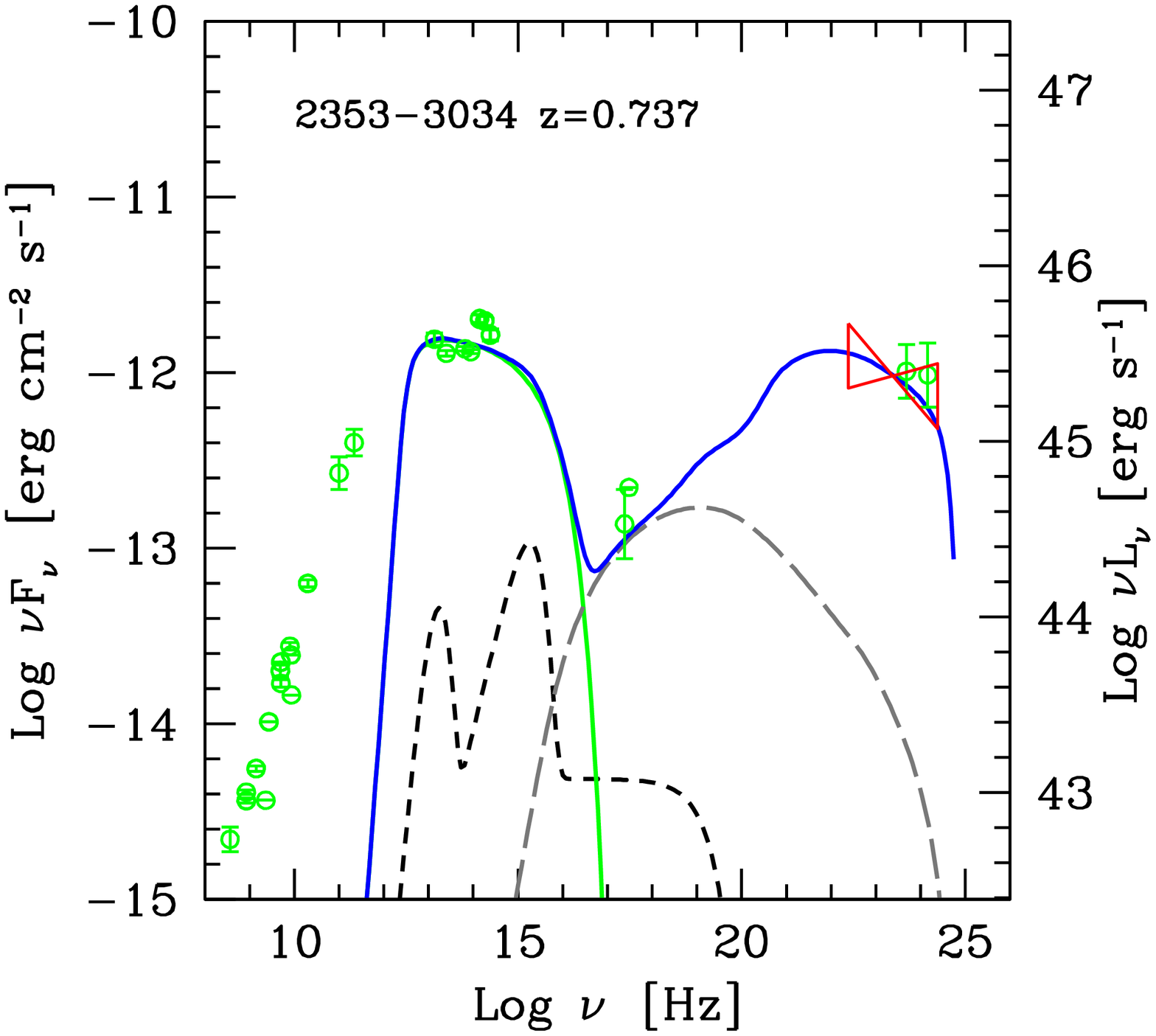,width=6cm,height=5.8cm}
\vskip -0.4 cm
\Large(a)
\vskip -0.1 cm
\hskip -0.2 cm
\psfig{figure=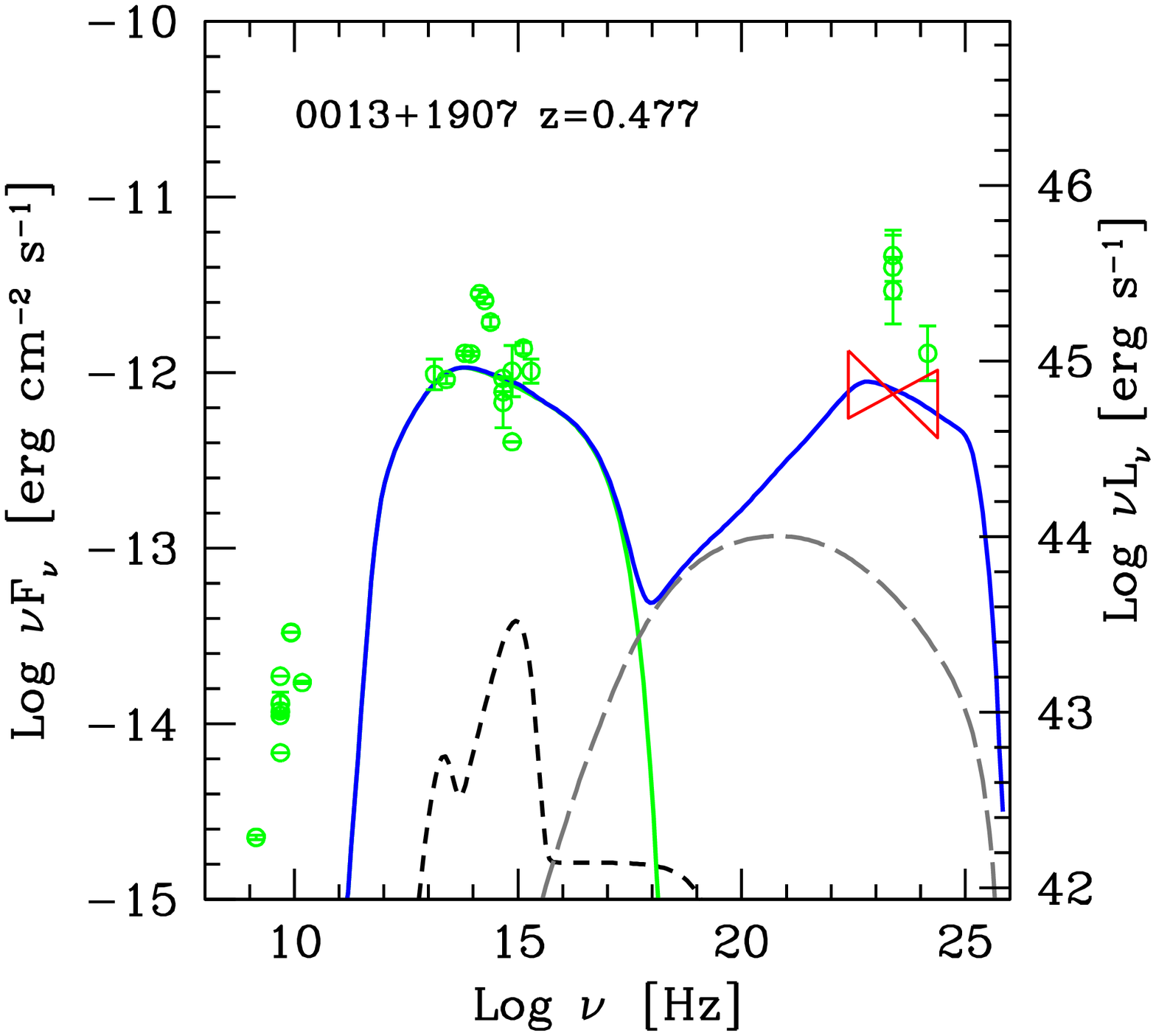,width=6cm,height=5.8cm}
\hskip -0.2 cm
\psfig{figure=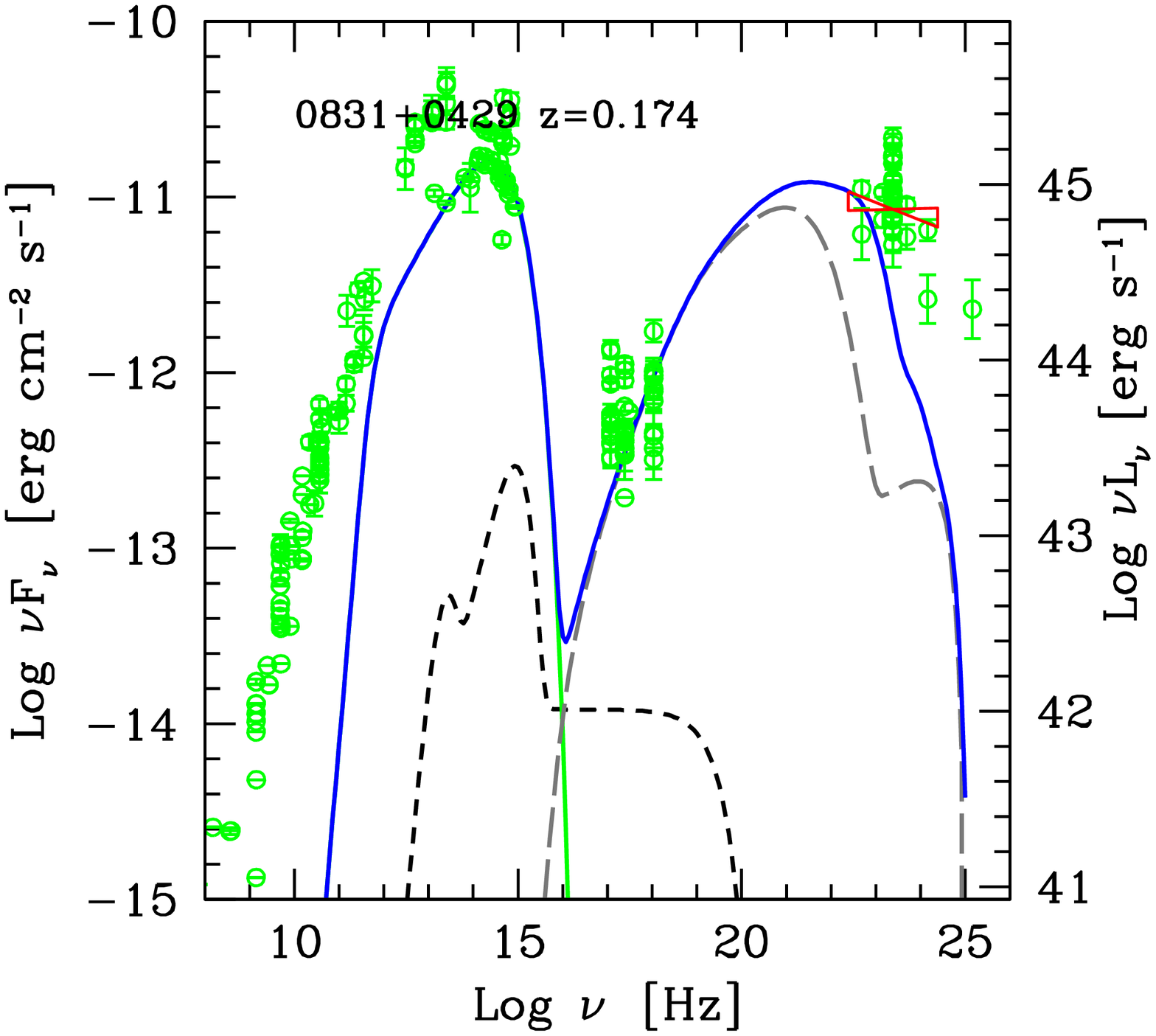,width=6cm,height=5.8cm}
\hskip -0.2 cm
\psfig{figure=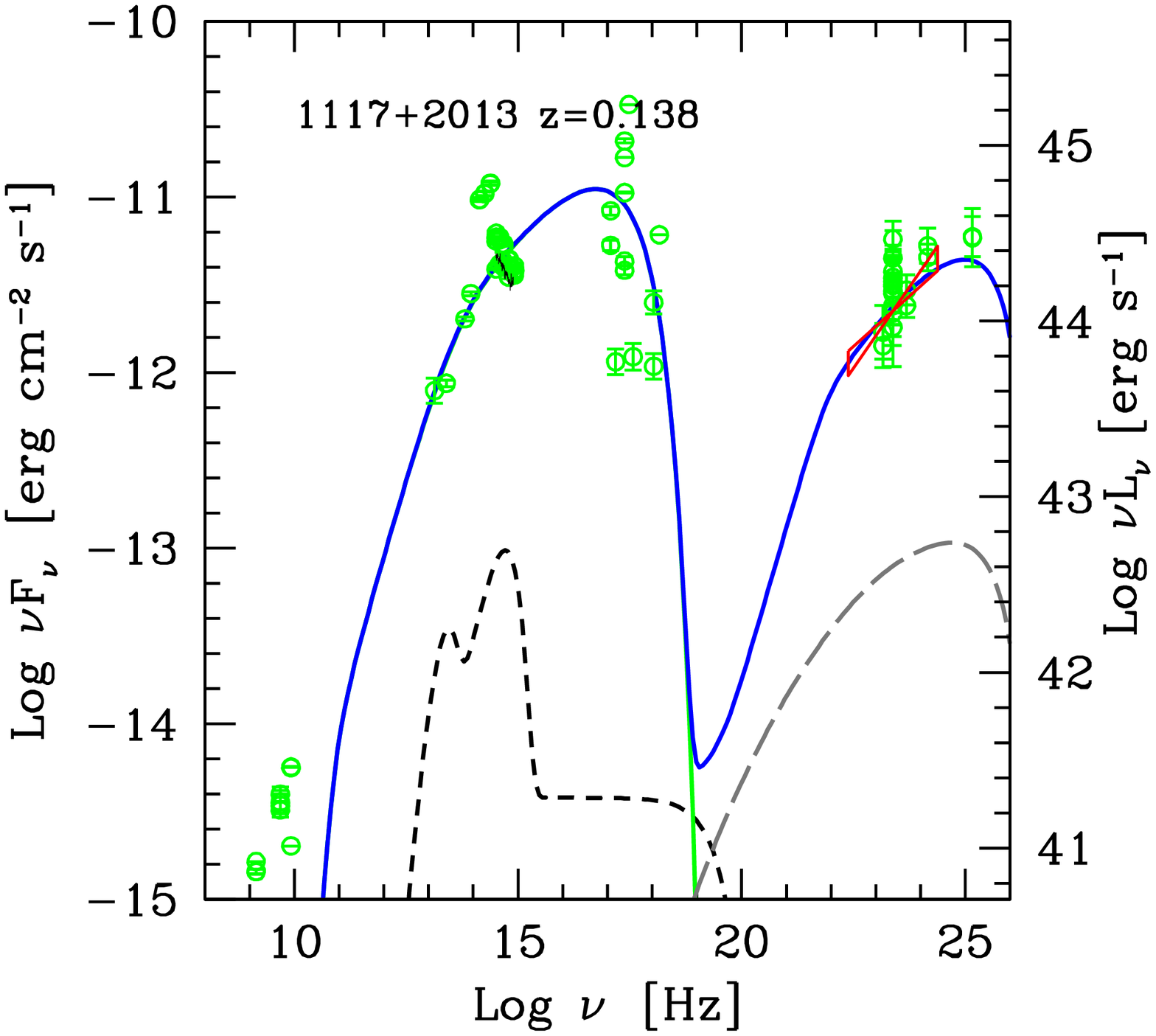,width=6cm,height=5.8cm}
\vskip -0.6 cm 
\hskip -0.2 cm
\psfig{figure=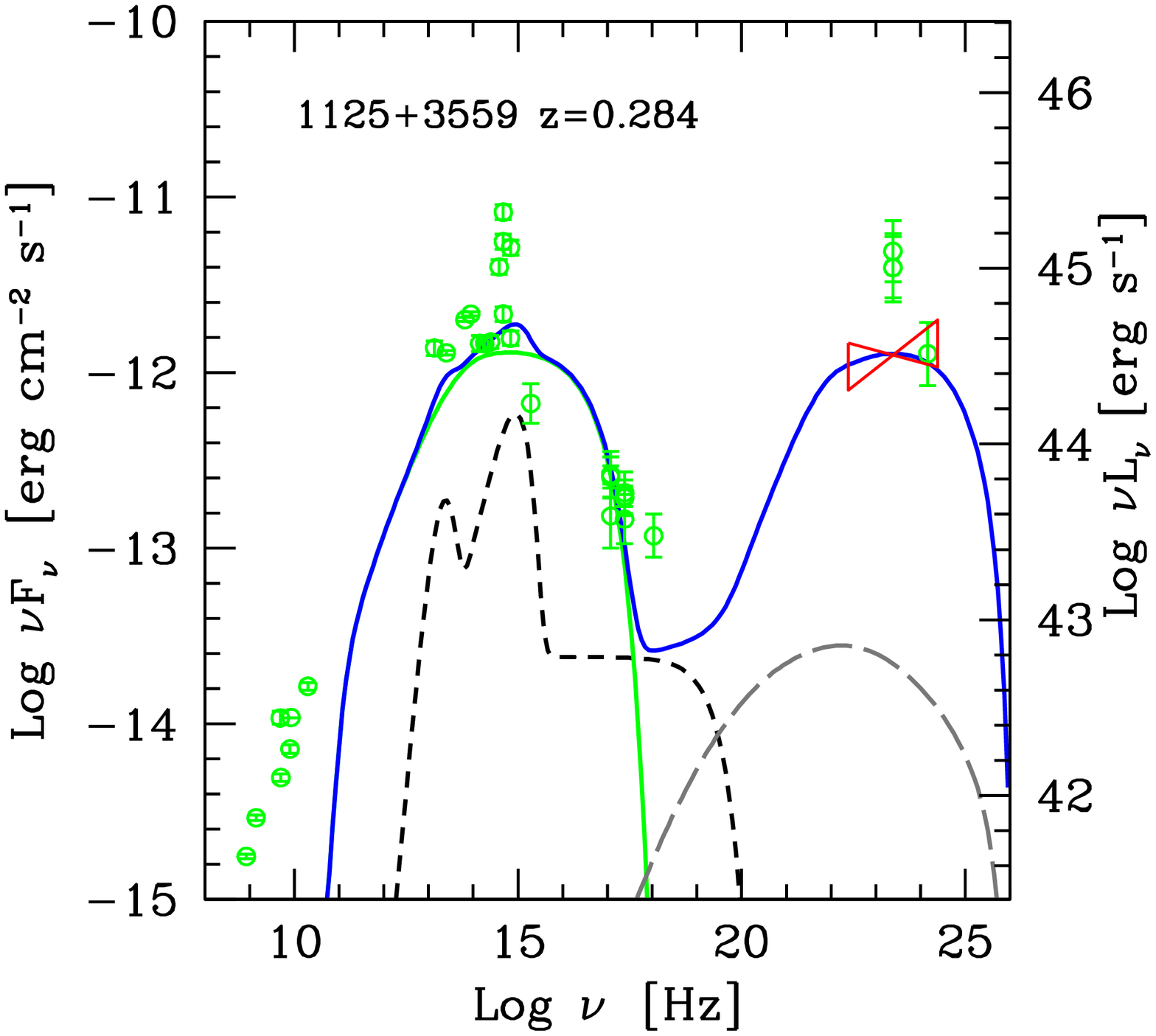,width=6cm,height=5.8cm}
\hskip -0.2 cm
\psfig{figure=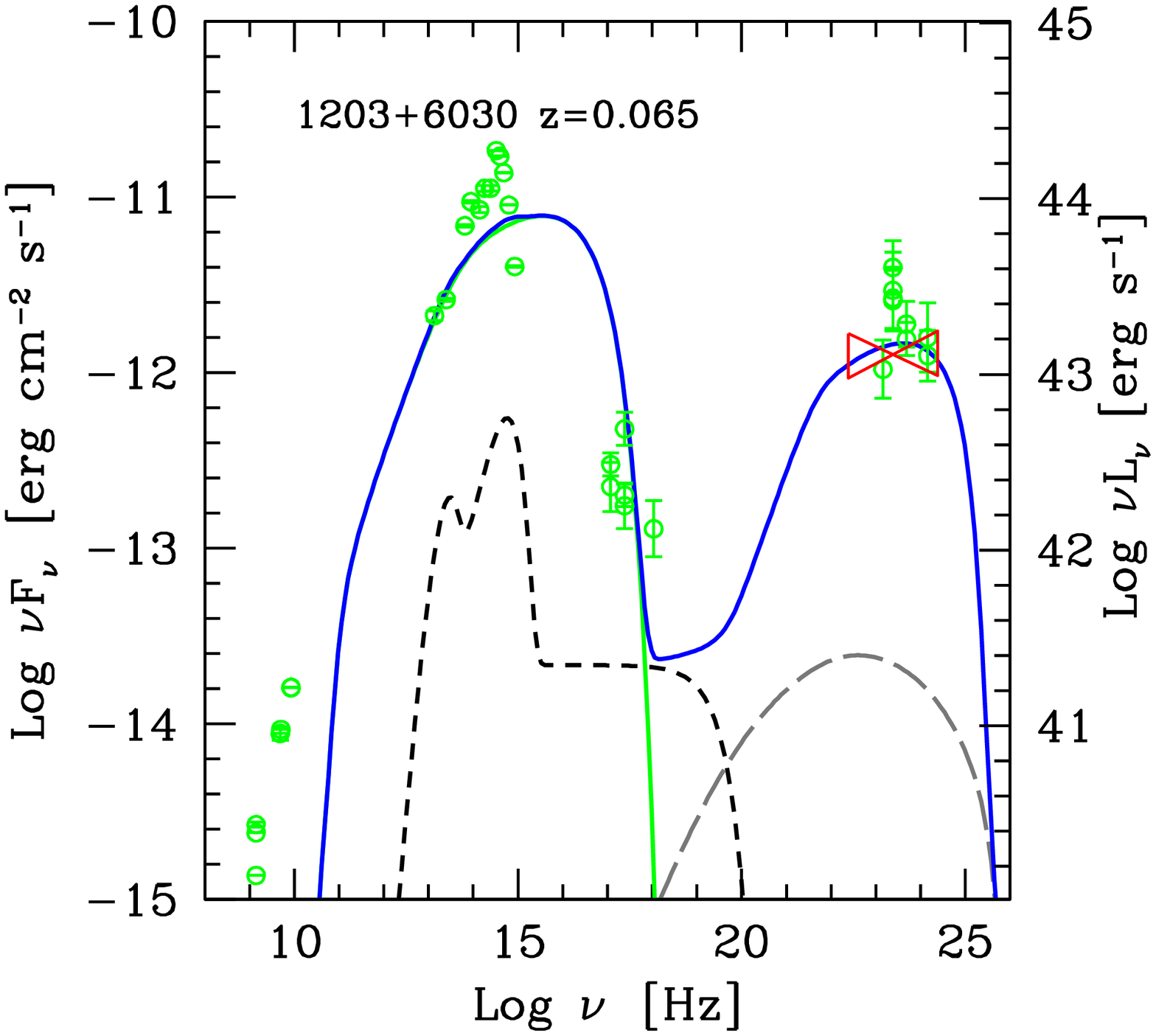,width=6cm,height=5.8cm}
\hskip -0.2 cm
\psfig{figure=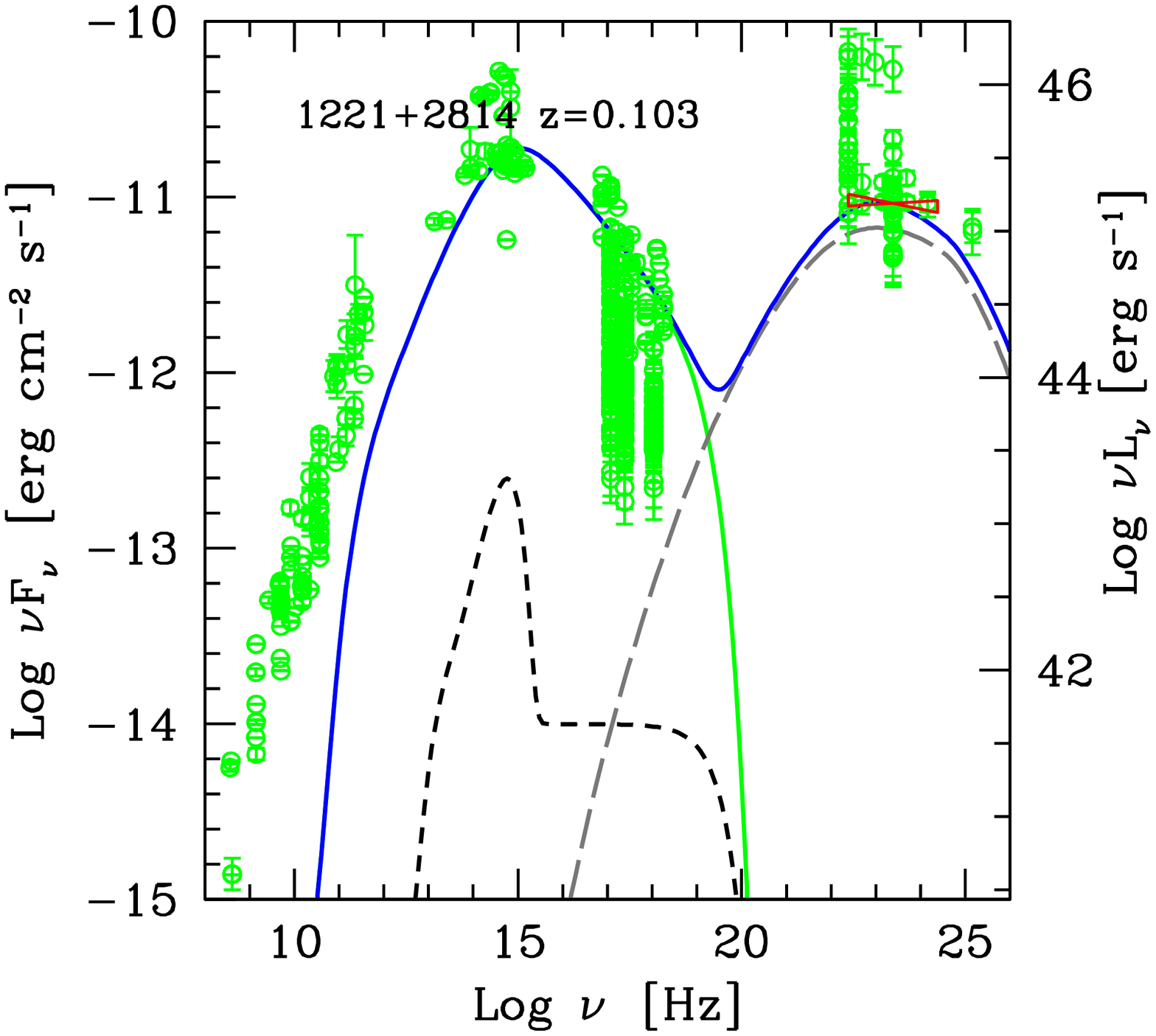,width=6cm,height=5.8cm}
\vskip -0.6 cm 
\hskip -0.2 cm
\psfig{figure=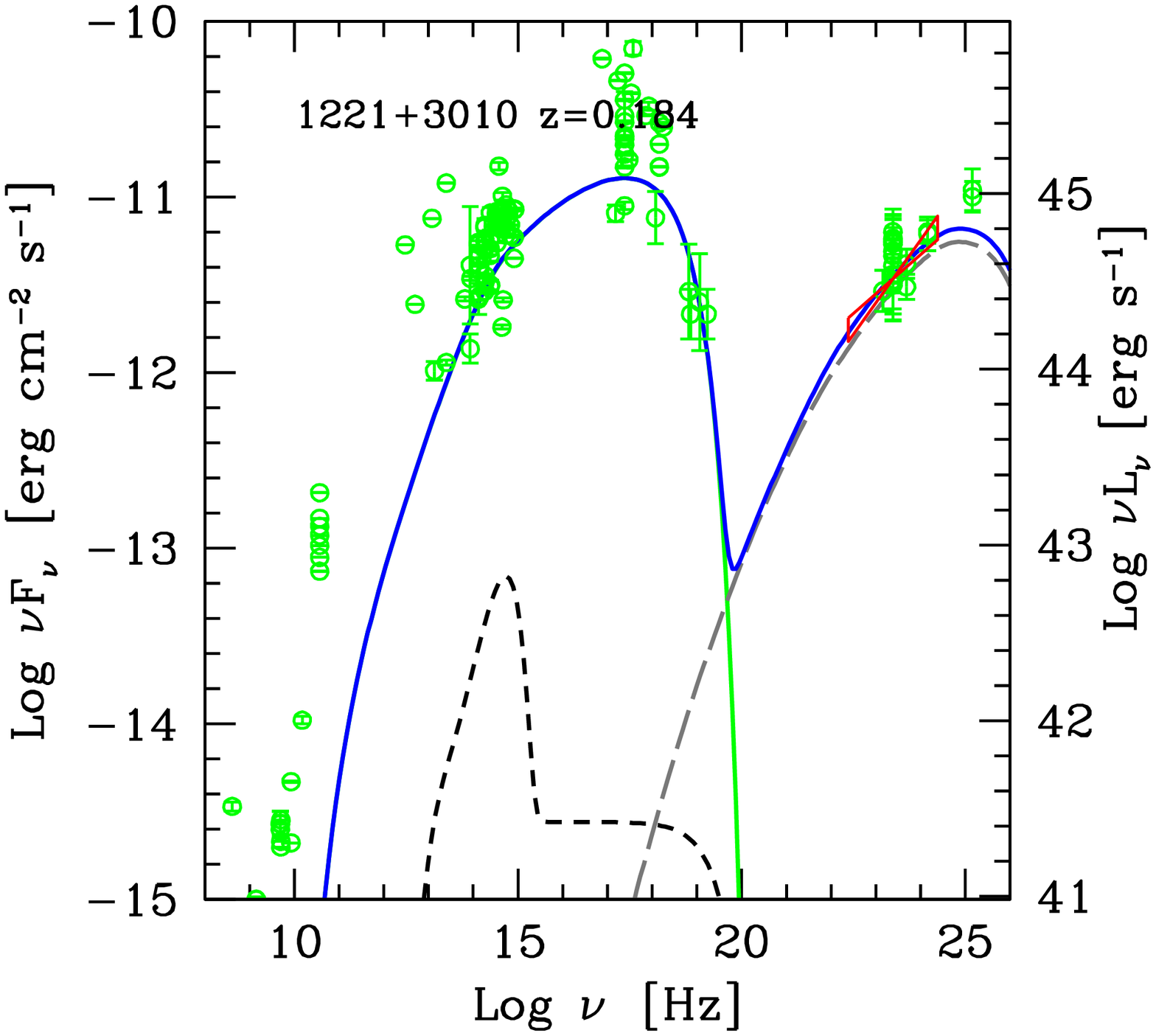,width=6cm,height=5.8cm}
\hskip -0.2 cm
\psfig{figure=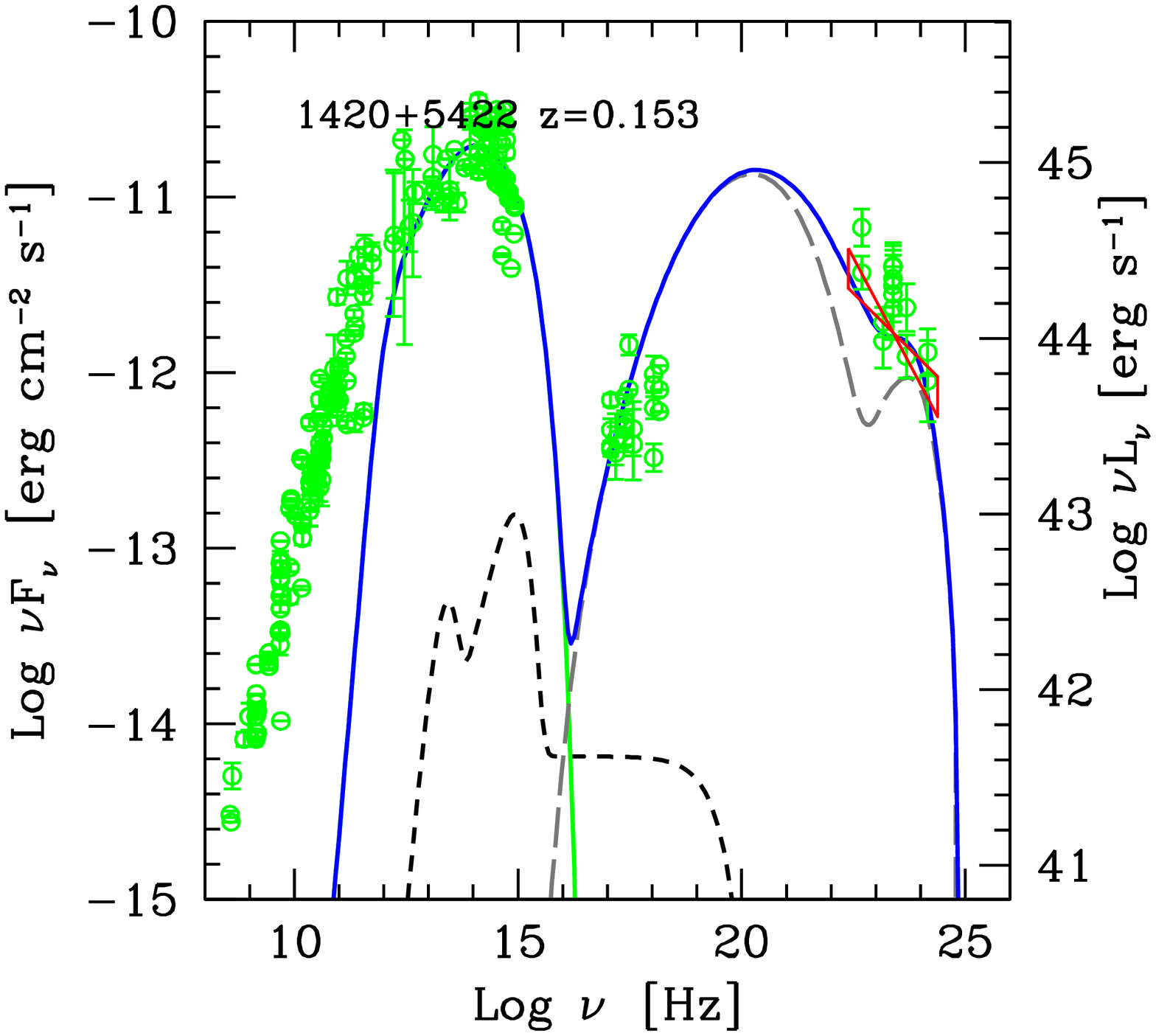,width=6cm,height=5.8cm}
\hskip -0.2 cm
\psfig{figure=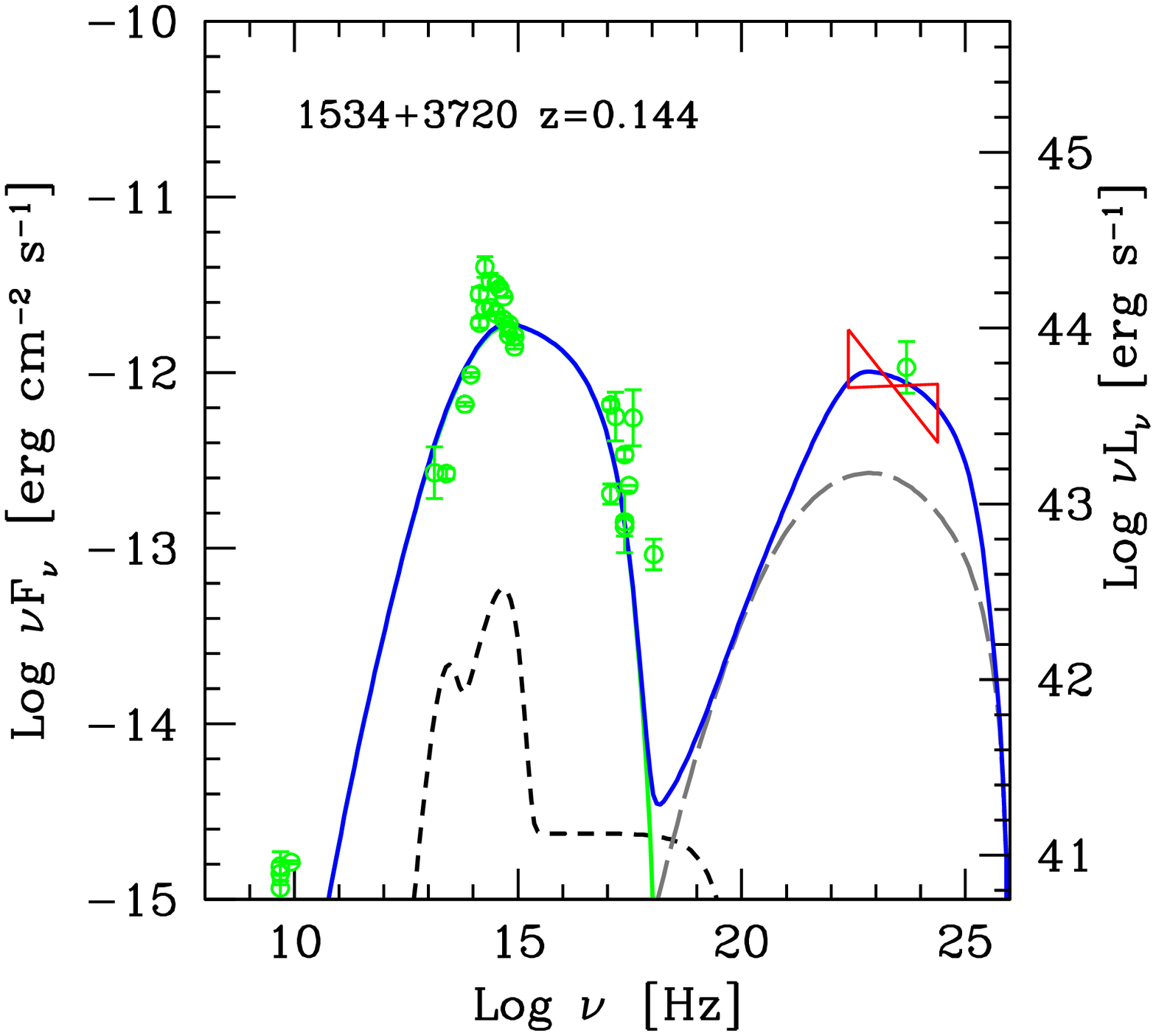,width=6cm,height=5.8cm}
\vskip -0.5 cm
(b)
\caption{(a) Sources reclassified as BL/FS.
(b) Sources that can be classified as BL Lacs, both 
from the optical spectra and from the SED aspect. 
Lines as in Figure \ref{sed_fs}.
}
\label{sed_bllac}
\end{figure*}

We show in Figures \ref{sed_fs}, \ref{sed_blfs} 
and \ref{sed_bllac} the SEDs of the 25 objects from S13, 
divided according to our reclassification, discussed in \S\ref{gamma}. 
The SEDs are fitted with a one--zone leptonic model, fully described 
in Ghisellini \& Tavecchio (2009).

\begin{thebibliography} {}

%
\bibitem[Abdo(2010)]{abd10a} Abdo A.A., Ackermann M., Ajello M. et al., 2010, ApJ, 715, 429  

%
%
\bibitem[Ackermann(2011)]{ack11} Ackermann M., Ajello M., Allafort A., et al., 2011, ApJ, 743, 171 

%
%
\bibitem[Baldwin(1978)]{bal78} Baldwin J.A \& Netzer H., 1978, ApJ, 226, 1

\bibitem[Baum(1989)]{bau89a} Baum S.A., \& Heckman, T., 1989a, ApJ, 336, 681

\bibitem[Baum(1989)]{bau89b} Baum, S.A., \& Heckman, T., 1989b, ApJ, 336, 702

\bibitem[Bettoni(2003)]{bet03} Bettoni D., Falomo R., Fasano G., Govoni F., 2003, A\&A, 399, 869

\bibitem[Buttiglione(2009)]{but09} Buttiglione S., Capetti A., Celotti A., et al., 2009, A\&A, 495, 1033

\bibitem[Buttiglione(2010)]{but10} Buttiglione S., Capetti A., Celotti A., et al., 2010, A\&A, 509, 6

\bibitem[Capetti(2010)]{cap10} Capetti A., Raiteri C.M., Buttiglione S., 2010, A\&A, 349, 77

\bibitem[Celotti(1997)]{cel97} Celotti A., Padovani P., \& Ghisellini G., 1997, MNRAS, 286, 415 %

\bibitem[Celotti(2008)]{cel08} Celotti A. \& Ghisellini G., 2008, MNRAS, 385, 283 %

%
%
%
%
\bibitem[Edwards(2002)]{edw02} Edwards P.G.\ \& Piner B.G., 2002, ApJ, 579, 67

\bibitem[Fanaroff(1974)]{fan74} Fanaroff B.L., Riley J.M, 1974, MNRAS, 167, 31

\bibitem[Francis(1991)]{fra91} Francis P.J., Hewett P.C., Foltz C.B., Chaffee F.H., Weymann R.J. %
             \& Morris S.L., 1991, ApJ, 373, 465   

\bibitem[Georganopoulos(2004)]{geo04} Georganopoulos M.\ \& Kazanas D., 2004, ApJ, 604, 81	

%
%
\bibitem[GG(2001)]{gg01} Ghisellini G. \& Celotti A., 2001, A\&A, 379, L1 

%
\bibitem[GG(2009)]{gg09} Ghisellini G. \& Tavecchio F., 2009, MNRAS, 397, 985  

\bibitem[GG(2005)]{gg05} Ghisellini G., Tavecchio F.\ \& Chiaberge M., 2005, A\&A, 432, 401

\bibitem[GG(2010)]{gg10} Ghisellini G., Tavecchio F., Foschini L., Ghirlanda G., Maraschi L. \& Celotti A., %
             2010, MNRAS, 402, 497  

\bibitem[GG(2011)]{gg11} Ghisellini G., Tavecchio F., Foschini L, Ghirlanda G., 2011, MNRAS, in press (G11) %

\bibitem[Giommi(2012)]{gio12} Giommi P., Padovani P., Polenta G., Turriziani S., 
		D'Elia V., Piranomonte S., 2012, MNRAS, 420, 2899

\bibitem[Giommi(2013)]{gio13} Giommi P., Padovani P., Polenta G., 2013, MNRAS, 431, 1914

\bibitem[Giovannini(1999)]{gio99} Giovannini G., et al., 1999, ApJ, 522, 101

\bibitem[Giroletti(2004)]{gir04} Giroletti M., Giovannini G., Feretti L., et al., 2004, ApJ, 600, 127 

\bibitem[Healey(2007)]{hea07} Healey S.E., et al.\ 2007, ApJS, 171, 61 

\bibitem[Komissarov(1990)]{kom90} Komissarov S.S., 1990, SvA, 16, L284

\bibitem[Labiano(2008)]{lab08} Labiano A., 2008, A\&A, 488, L59

%
\bibitem[Landt(2004)]{lan04} Landt H., Padovani P., Giommi P. \ Perlman E.S., 2004, MNRAS, 351, 83 

\bibitem[Laing(1993)]{lai93} Laing R., 1993, in Burgarella D., Livio M., OÕDea C., eds, 
		Astrophysical Jets. Cambridge Univ.\ Press, Cambridge, p. 95

\bibitem[Laing(1994)]{lai94} Laing R.A., Jenkins C.R., Wall J.V., \& Unger S.W., 1994, 
		in The Physics of Active Galaxies, ed. G. V. Bicknell, 
		M.A.\ Dopita, \& P.J.\ Quinn, ASP Conf.\ Ser., 54, 201

%
\bibitem[Mahadevan(1997)]{mah97} Mahadevan R., 1997, ApJ, 447, 585 %

%
\bibitem[Marconi(2003)]{mar03} Marconi A.\ \& Hunt L.K. 2003, ApJ, 589, L21

%
\bibitem[Morganti(1997)]{mor97} Morganti R., Tadhunter C.N., Dickson R., \& Shaw M., 1997, A\&A, 326, 130

\bibitem[Narayan(1997)]{nar97} Narayan R., Garcia M.R. \& McClintock J.E., 1997, ApJ, 478, L79 %

\bibitem[Narayan(1995)]{nar95} Narayan R. \& Yi I., 1995, ApJ, 452, 710 %

\bibitem[Owen(1989)]{owe89} Owen F.N., Hardee P.E.\ \& Cornwell T.J., 1989, ApJ, 340, 698

%
\bibitem[PP(1995)]{pp95} Padovani P.\ \& Giommi P., 1995,  ApJ, 444, 567 

\bibitem[Piner(2004)]{pin04} Piner B.G.\ \& Edwards P.G., 2004, 600, 115

%
%
\bibitem[Plotkin(2011)]{plo11} Plotkin R.M., Markoff S., Trager S.C. \& Anderson S.F., 2011, MNRAS, 413, 805 

\bibitem[Rawlings(1991)]{raw91} Rawlings S., \& Saunders R., 1991, Nature, 349, 138

\bibitem[Rawlings(1989)]{raw89} Rawlings S., Saunders R., Eales S.A., \& Mackay C.D., 1989, MNRAS, 240, 701

\bibitem[TS(2012)]{ts12a} Sbarrato T., Ghisellini G., Maraschi L., Colpi M., 2012, MNRAS, 421, 1764 (TS12)

\bibitem[Sbarufatti(2005)]{sba05} Sbarufatti B., Treves A.\ \& Falomo R., 2005, ApJ, 635, 173

%
\bibitem[Shakura(1973)]{sha73} Shakura N.I. \& Sunyaev R.A., 1973 A\&A, 24, 337    %

\bibitem[Sharma(2007)]{sha07} Sharma P., Quataert E., Hammet G.H.\ \& Stone J.M., 2007, ApJ, 667, 714 %

\bibitem[Shaw(2012)]{sha12} Shaw M.S., et al., 2012, ApJ, 748, 49 (S12)

\bibitem[Shaw(2013)]{sha13} Shaw M.S., et al., 2013, ApJ, 764, 135 (S13)

\bibitem[Shen(2011)]{she11} Shen Y., Richards G.T., Strauss M.A. et al., 2011, ApJS, 194, 45 

\bibitem[Smith(1981)]{smi81} Smith M.G., Carswell R.F., Whelan J.A.J. et al., 1981, MNRAS 195, 437 

\bibitem[Spinrad(1985)]{spi85} Spinrad H., Marr J., Aguilar L.\ \& Djorgovski S. 1985, PASP, 97, 932 

\bibitem[Swain(1998)]{swa98} Swain M.R., Bridle A.H., Baum S.A., 1998, ApJ, 507, 29

%
\bibitem[UP(1995)]{up95} Urry C.M.\ \& Padovani P., 1995, PASP, 107, 803 %

\bibitem[Vermeulen(1995)]{ver95} Vermeulen R.C., et al., 
		2005, ApJ, 452, L5

%
\bibitem[Willott(1999)]{wil99} Willott C.J., Rawlings S., Blundell K.M., \& Lacy M., 1999, MNRAS, 309,1017



\end{thebibliography}
\end{document}